\documentclass[journal]{IEEEtran}

\usepackage{cite}
\usepackage{graphicx}
\usepackage[lowtilde]{url}
\usepackage[caption=false,font=footnotesize]{subfig}
\usepackage{pgfplots}
\pgfplotsset{compat=newest}
\usetikzlibrary{plotmarks}
\usepackage{grffile}
\usepackage{mathtools}
\usepackage{amssymb}
\usepackage{amsmath}
\usepackage{multirow}
\usepackage{enumitem}
\usepackage[euler]{textgreek}
\usepackage{booktabs} 
\usepackage{array}
\usepackage[ruled, linesnumbered]{algorithm2e}

\makeatletter
\long\def\@makecaption#1#2{\ifx\@captype\@IEEEtablestring%
\footnotesize\begin{center}{\normalfont\footnotesize #1}\\
{\normalfont\footnotesize\scshape #2}\end{center}%
\@IEEEtablecaptionsepspace
\else
\@IEEEfigurecaptionsepspace
\setbox\@tempboxa\hbox{\normalfont\footnotesize {#1.}~~ #2}%
\ifdim \wd\@tempboxa >\hsize%
\setbox\@tempboxa\hbox{\normalfont\footnotesize {#1.}~~ }%
\parbox[t]{\hsize}{\normalfont\footnotesize \noindent\unhbox\@tempboxa#2}%
\else
\hbox to\hsize{\normalfont\footnotesize\hfil\box\@tempboxa\hfil}\fi\fi}
\makeatother

\newcommand{\RN}[1]{%
  \textup{\uppercase\expandafter{\romannumeral#1}}%
}

\makeatletter
\newcommand{\removelatexerror}{\let\@latex@error\@gobble}
\makeatother

\DeclareMathOperator*{\argmin}{arg\,min}

\newcolumntype{C}[1]{>{\centering\let\newline\\\arraybackslash\hspace{0pt}}m{#1}}

\begin{document}
\title{Characterizing Generalized Rate-Distortion Performance of Video Coding:\\ An Eigen Analysis Approach}
\author{Zhengfang~Duanmu*,~\IEEEmembership{Student Member,~IEEE,}
        Wentao~Liu*,~\IEEEmembership{Student Member,~IEEE,}\\
        Zhuoran~Li,~\IEEEmembership{Student Member,~IEEE,}
        Kede~Ma,~\IEEEmembership{Member,~IEEE,}
        and~Zhou~Wang,~\IEEEmembership{Fellow,~IEEE}
\thanks{Zhengfang~Duanmu, Wentao~Liu, Zhuoran~Li, and Zhou~Wang are with the Department of Electrical and Computer Engineering, University of Waterloo, Waterloo, ON N2L 3G1, Canada (e-mail: \{zduanmu, w238liu, z777li, zhou.wang\}@uwaterloo.ca). *Authors contributed equally.}
\thanks{Kede Ma is with the Department of Computer Science,  City University of Hong Kong, Kowloon, Hong Kong (e-mail: kede.ma@cityu.edu.hk).}
}

\maketitle

\begin{abstract}
Rate-distortion (RD) theory is at the heart of lossy data compression. Here we aim to model the generalized RD (GRD) trade-off between the visual quality of a compressed video and its encoding profiles (\textit{e.g.}, bitrate and spatial resolution).
We first define the theoretical functional space $\mathcal{W}$ of the GRD function  by analyzing its mathematical properties.
We show that $\mathcal{W}$ is a convex set in a Hilbert space, inspiring a computational model of the GRD function, and a method of estimating model parameters from sparse measurements. 
To demonstrate the feasibility of our idea, we collect a large-scale database of real-world GRD functions, which turn out to live in a low-dimensional subspace of $\mathcal{W}$.
Combining the GRD reconstruction framework and the learned low-dimensional space, we create a low-parameter eigen GRD method to accurately estimate the GRD function of a source video content from only a few queries.
Experimental results on the database show that the learned GRD method significantly outperforms state-of-the-art empirical RD estimation methods both in accuracy and efficiency. 
Last, we demonstrate the promise of the proposed model in video codec comparison.
\end{abstract}


\IEEEpeerreviewmaketitle

\section{Introduction}\label{sec:intro}

\IEEEPARstart{R}{ate-distortion} (RD) theory lays a theoretical foundation for lossy data compression, and is widely used to guide the design of image and video compression schemes~\cite{berger1975rate}.
One of the most profound outcomes from the theory is the so-called RD function~\cite{shannon1959coding}, which describes the minimum bitrate required to encode a signal when a fixed amount of distortion is allowed (\textit{i.e.}, the highest achievable quality given limited bitrate resources).
Many multimedia applications depend on precise measurements of RD functions to characterize source videos, to maximize user Quality-of-Experience (QoE), and to make efficient use of bitrate resources.
Examples of such applications include codec evaluation~\cite{grois2013performance,bitmovin2018codec}, rate-distortion optimization~\cite{wang2012ssim}, video quality assessment (VQA)~\cite{ou2014q}, encoding representation recommendation~\cite{zhang2013qoe,toni2015optimal,de2016complexity,chen2016subjective}, and QoE optimization of streaming videos~\cite{wang2015objective,chen2017encoding}.

Despite the tremendous growth in multimedia applications over the years, effective methods for estimating RD functions are largely lacking.
Previous works~\cite{wang2012ssim,chen2016subjective,toni2015optimal,bjontegaard2001bda,bjontegaard2008bdb} mainly focused on one-dimensional RD function estimation while holding other video attributes fixed~\cite{duanmu2019modeling}. 
For example, Toni~\textit{et al.}~\cite{toni2015optimal,kreuzberger2016comparative} assumed a reciprocal parametric form for the RD function. 
Chen {\it et al.}~\cite{chen2016subjective} modeled the rate-quality curve at each spatial resolution using a logarithmic function. 
The above methods make strong \textit{a priori} assumptions, which may not hold in real-world situations. 
Later, relaxed constraints on the RD function such as continuity~\cite{de2016complexity} and axial monotonicity~\cite{duanmu2019modeling} are imposed. 
However, such RD function estimation methods often require dense sampling in the video representation space, which is computationally expensive.

\begin{figure*}
	\centering
	\includegraphics[width=\textwidth]{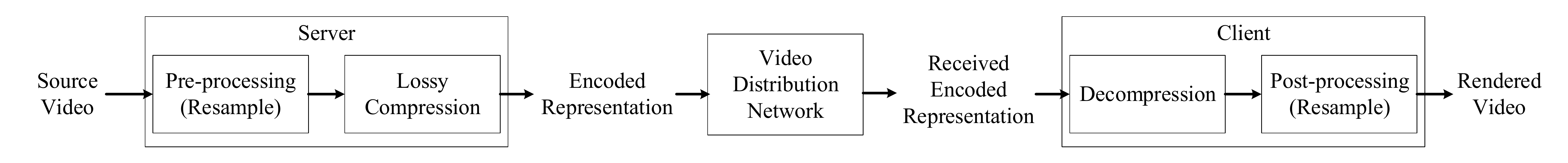}
	\caption{Flow diagram of visual communication pipeline.}\label{fig:pipeline}
\end{figure*}

However, bitrate is not the only influential factor of perceptual video quality.
In order to meet the growing diversities of display technology, video content, and network capacity, practical video delivery is accomplished by the cooperation of two components: the server and the client. 
Fig.~\ref{fig:pipeline} shows the visual communication pipeline.
At the server side, source videos are pre-processed and encoded into several representations with different bitrates, spatial resolutions, frame rates, and bit depths to fit different communication channels.
At the client side, video players resample the transmitted videos to fit the displays~\cite{ISO2012Dash}.
The perceived video quality is altered by all these distortion processes and the interactions among them.
Computational models that can predict such end-of-process video quality given the set of encoding parameters are of great interest in the video processing community~\cite{de2016complexity,toni2015optimal,zhang2013qoe,chen2017encoding}.

To this end, we formalize the concept of generalized rate-distortion (GRD) function, and construct mathematical and computational models to characterize it for compressed videos.
In this paper, we define the GRD function by $f: \mathbb{R}^2 \mapsto \mathbb{R}$, where the input is the encoding bitrate and the spatial resolution, and the output is the video quality. 
A key feature of the GRD function is that it is content- and encoder-dependent.

While many recent studies acknowledge the importance of the GRD function~\cite{zhang2013qoe,toni2015optimal,de2016complexity,chen2016subjective}, existing GRD models often rely on heuristically designed functional forms, without any theoretical justification or empirical validation. 
Moreover, a recent study in GRD functions has shown that the RD curves at different resolutions of a source video are highly dependent~\cite{duanmu2019modeling}.
Most existing methods estimate RD functions with different video attributes in an independent manner~\cite{netflix2015pertitle,zhang2013qoe,wang2015objective}, completely ignoring the regularization among such RD functions, an essential property of GRD functions~\cite{duanmu2019modeling}.
In addition, their performance is sensitive to the number of attribute-quality pairs for training, and degrades drastically when the sampled pairs are sparse. 
This scenario often occurs in practice because obtaining an attribute-quality pair involves sophisticated video encoding and quality assessment processes, both of which may demand excessive computational resources. 
For example, the recently announced AV1 video encoder~\cite{aom2018AV1} can be over 100 times slower than real-time for full high-definition (\textit{i.e.}, $1920\times 1080$) video content~\cite{li2019avc,ozer2019good}. 

We believe the major difficulty arises from the lack of thorough theoretical understanding and accurate computational modeling of GRD functions. 
Inspired by previous work in modeling camera response function~\cite{grossberg2004modeling}, we perform mathematical analysis of GRD functions, based on which we describe a computational model for accurate GRD function reconstruction, whose desirable properties are as follows:
\begin{itemize}
  \item Mathematical soundness. 
  We analyze the mathematical properties that all GRD functions share, and show that they must lie within a convex set $\mathcal{W}$ resulting from the intersection of an affine subspace and a convex cone in a Hilbert space. 
  This analysis not only inspires a computational GRD model, but also guarantees the validity of the estimated GRD function. 

  \item Low complexity. 
  We collect a great number of real-world GRD functions, and find that they live in a low-dimensional subspace of $\mathcal{W}$, suggesting efficient model estimation with a minimal number of samples. 

  \item Quality. 
  We conduct extensive experiments to show that the proposed method achieves consistent improvement both in prediction accuracy and rate of convergence. 
  The robustness of the proposed method is also empirically validated in various practical scenarios.
\end{itemize}
In addition, we demonstrate how video codec comparison can benefit from the proposed GRD model.
We have made the proposed GRD model along with the GRD function database available at \url{http://ece.uwaterloo.ca/~w238liu/2020egrd/}.

\section{Theoretical Space of GRD Functions}\label{sec:theoretic}
\begin{figure*}[t]
    \centering
	\subfloat[]{\includegraphics[width=0.33\textwidth]{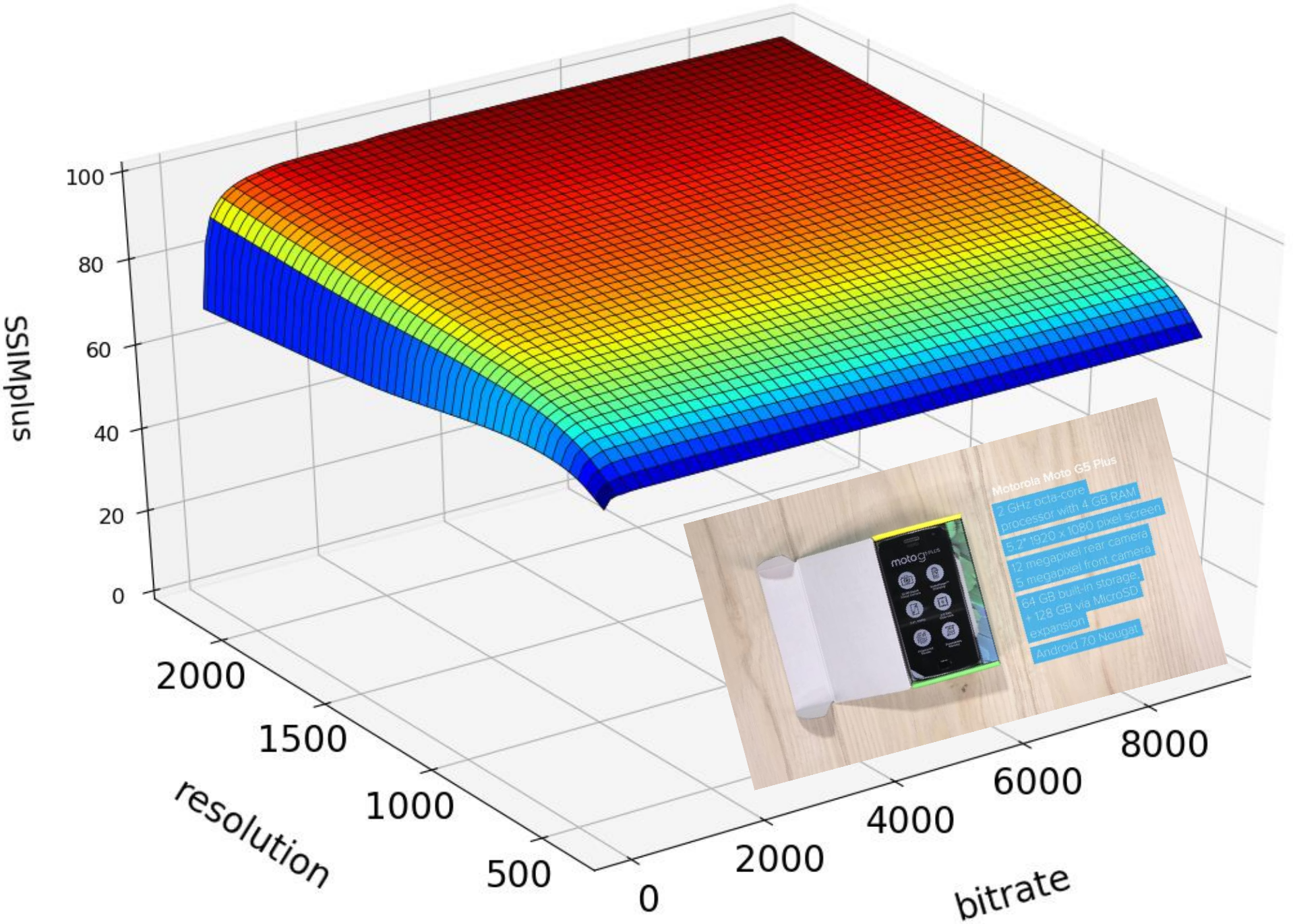}}
    \subfloat[]{\includegraphics[width=0.33\textwidth]{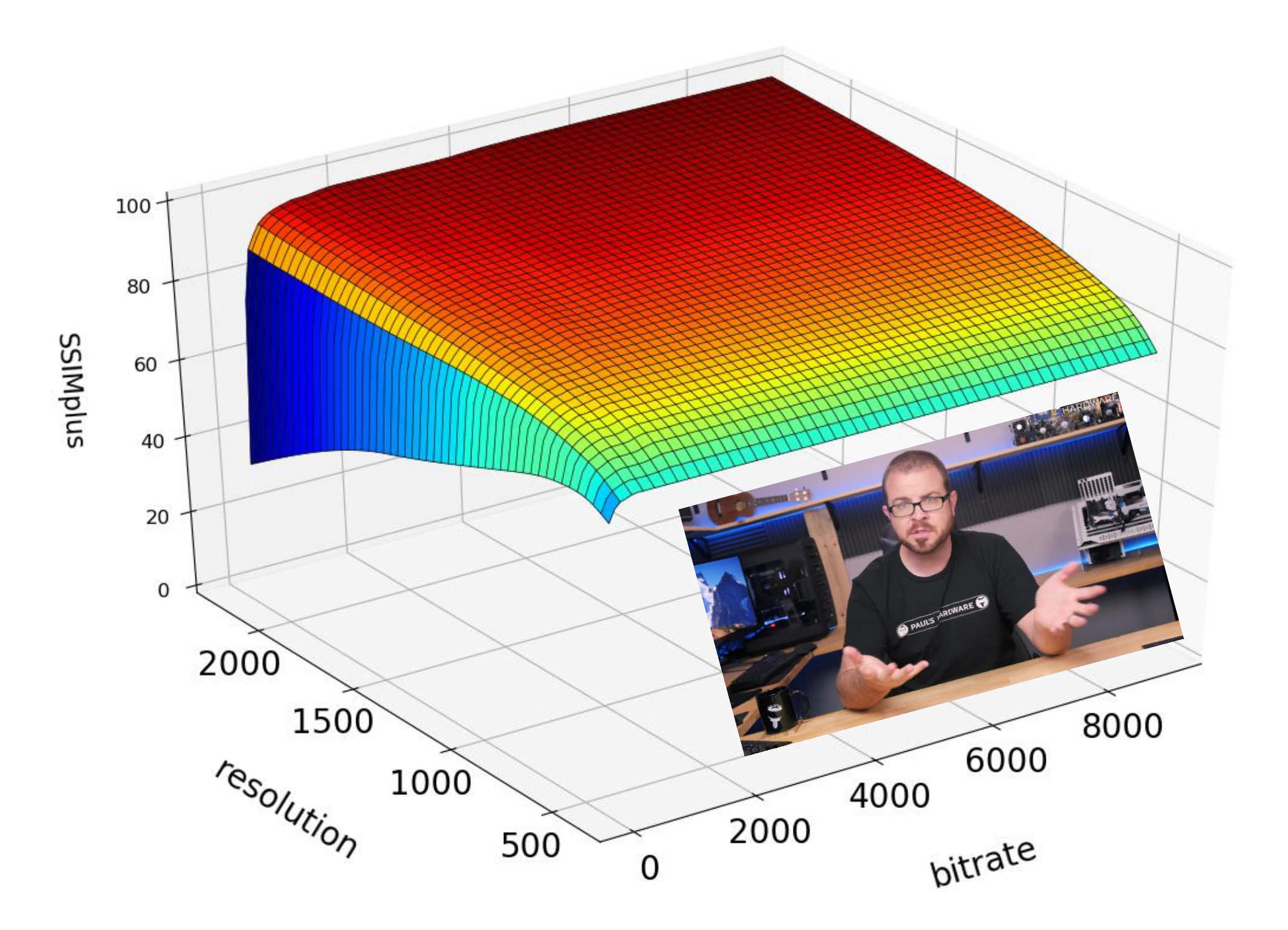}}
    \subfloat[]{\includegraphics[width=0.33\textwidth]{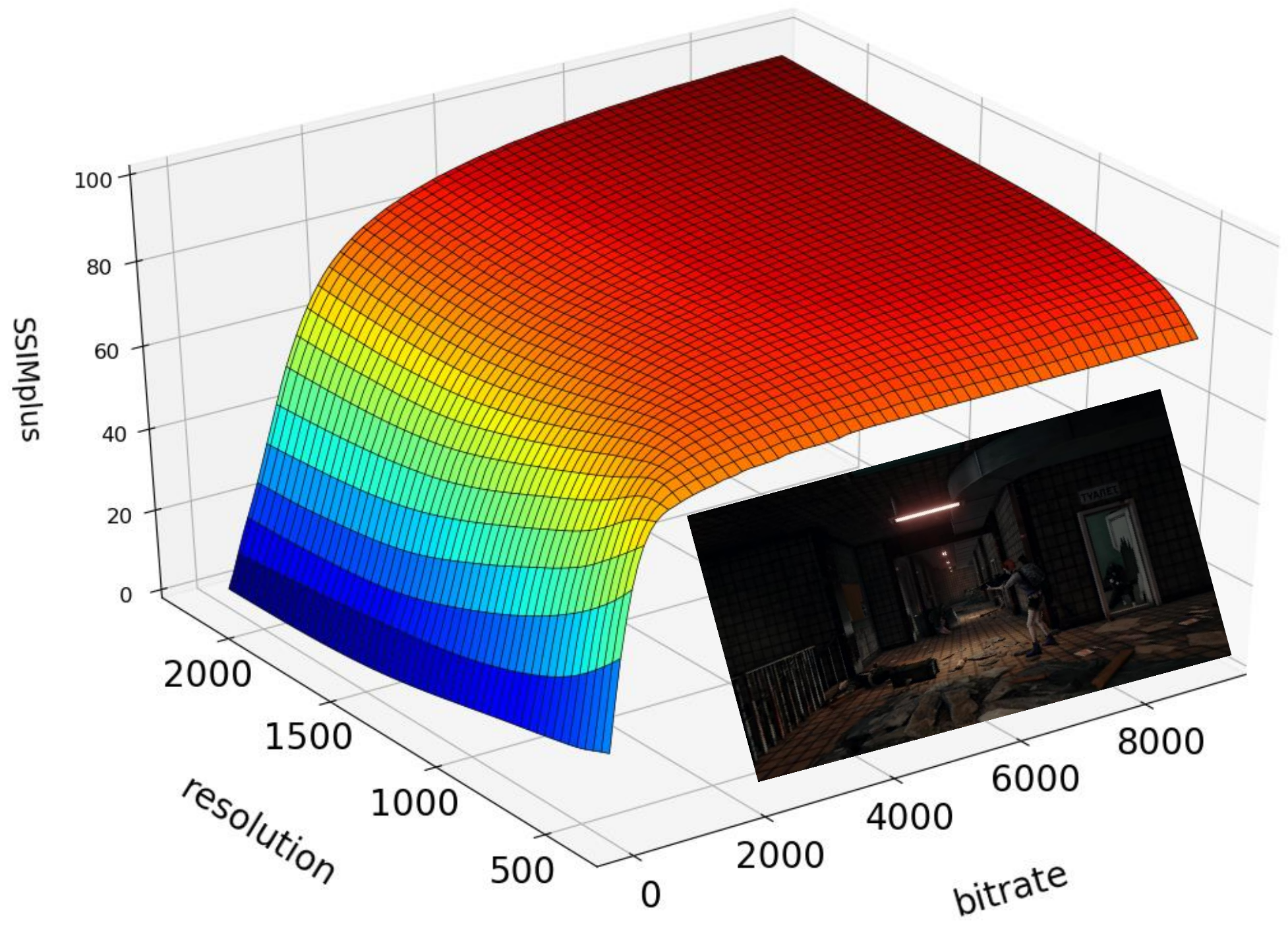}}
    \caption{Samples of GRD surfaces for different video content.}\label{fig:sample_grd}
\end{figure*}

We begin by stating our assumptions of desirable GRD functions. 
Our first assumption is that the domain of GRD functions is a compact set $\Omega$.
A typical setting of $\Omega$ is a rectangular region in the bitrate-resolution space, \textit{i.e.}, $(x, y) \in \Omega = [x_{\min}, x_{\max}] \times [y_{\min}, y_{\max}]$, where $x$ and $y$ represent the bitrate and diagonal length of spatial resolution of an encoded video representation, respectively.
The upper and lower bounds of $x$ and $y$ are easily determined in practical applications.
In this paper, we consider $x_{\min} = 0$, suggesting that all pixel intensities are severely degraded to a single value and therefore no bits are required to encode the video. 
This further implies that $\forall y \in [y_{\min}, y_{\max}], f(0, y) = z_{\min}$, where $z_{\min}$ represents the worst perceptual quality. 
The value of $x_{\max}$ may be determined by the maximum lossless encoding rate among a large number of pristine videos of diverse complexity. 
On the other hand, $y_{\max}$ is typically equal to the size of the source video, and $y_{\min}$ can be obtained from the commonly used encoding configuration recommendations~\cite{netflix2015pertitle,applea,grafl2013combined}.
In addition, since the unit of perceptual quality is arbitrary, we normalize the range of GRD functions such that $z_{\min}=0$ and $z_{\max}=100$~\cite{bt500subjective,sheikh2006statistical}.

Our second assumption is that GRD functions are continuous, \textit{i.e.}, $f \in C(\Omega)$.
In principle, RD curves are guaranteed to be continuous at each single resolution~\cite{berger1975rate}.
Besides, successive changes in spatial resolution would gradually deviate the spectrum of the source video, leading to smooth transitions in perceptual quality. The continuity of the GRD function has been empirically observed in many subjective user studies~\cite{ou2014q,zhai2008cross}.

Our third assumption is that GRD functions are axially monotonic along the bitrate dimension\footnote{In this work, we use rate-distortion function and rate-quality function interchangeably. Without loss of generality, we assume the function $f$ to be axially monotonically increasing. 
}. 
According to the RD theory~\cite{berger1975rate}, the perceived quality of the source video increases monotonically with respect to the number of bits it takes in lossy compression. 
However, such monotonicity constraint may not be applied to the spatial resolution dimension.
For example, encoding at high resolution with insufficient bitrate would produce artifacts such as blocking, ringing, and contouring, whereas encoding at low resolution with upsampling/interpolation using the same bitrate would introduce blurring. 
The relative quality resulting from the two encoding profiles is highly dependent on  the video content and the bitrate used.
Consequently, encoding at high spatial resolution may even result in lower video quality than encoding at low spatial resolution under the same bitrate~\cite{de2016complexity}. Fig.~\ref{fig:sample_grd} visualizes a few sample GRD surfaces to show the axial monotonicity and the continuity of real-world GRD functions.

Our fourth assumption is that GRD functions are monotonically increasing with respect to the spatial resolution at the highest encoding bitrate $x_{\max}$. 
When a pristine video is encoded with the highest bitrate, we consider that no compression artifacts will be introduced during encoding.
Therefore, quality degradation can only result from the loss of high frequency component during the lowpass filtering, downsampling and upsampling process.
Since the degree of frequency loss is a monotonic function of the scaling factor, the perceptual quality degrades as the encoding resolution reduces. 
This also implies that ($x_{\max}$, $y_{\max}$) corresponds to the highest perceptual quality $z_{\max}$.

Under these assumptions, we define the space of GRD functions as:
\begin{align}
\nonumber\mathcal{W} \coloneqq& \{f : \mathbb{R}^2 \mapsto \mathbb{R}| f \in C(\Omega); f(x_{\max}, y_{\max})=100; \\
\nonumber & f(x_{\min}, y)=0, \forall y \in [y_{\min}, y_{\max}]; \\
\nonumber & f(x_a, y) \leq f(x_b, y), \forall x_a \leq x_b; \\
& \text{and }f(x_{\max}, y_a) \leq f(x_{\max}, y_b), \forall y_a \leq y_b\}. \label{eq:grd_space}
\end{align}
The equality constraints in $\mathcal{W}$ jointly form an affine space $\mathcal{H}_1$, which can be described as a linear subspace
\begin{align}
\nonumber \mathcal{H}_0 \coloneqq &\{f : \mathbb{R}^2 \mapsto \mathbb{R}| f \in C(\Omega); f(x_{\max}, y_{\max})=0 \\
&  \mathrm{and \ } f(x_{\min}, y)=0, \forall y \in [y_{\min}, y_{\max}]\} \label{eq:grd_h0}
\end{align}
translated by any function $f_0\in \mathcal{H}_1$. 
Formally, we may express the relationship between $\mathcal{H}_1$ and $\mathcal{H}_0$ by
\begin{equation}
    \mathcal{H}_1 = f_0 + \mathcal{H}_0, \forall f_0 \in \mathcal{H}_1. \label{eq:grd_h1}
\end{equation}
The remaining inequality constraints jointly form a convex cone
\begin{align}
\nonumber \mathcal{V} \coloneqq &\{f : \mathbb{R}^2 \mapsto \mathbb{R}| f(x_a, y) \leq f(x_b, y), \forall x_a < x_b \\
& \mathrm{and \ } f(x_{\max}, y_a) \leq f(x_{\max}, y_b), \forall y_a < y_b\}, \label{eq:grd_cone}
\end{align}
as it is readily shown that $\forall \alpha, \beta \geq 0$ and $v_0, v_1 \in \mathcal{V}$, $\alpha v_0 + \beta v_1 \in \mathcal{V}$.

Finally, we conclude that the theoretical space $\mathcal{W}$ can be described as the intersection of the affine space $\mathcal{H}_1$ and the convex cone $\mathcal{V}$:
\begin{equation}\label{eq:theoretic}
    \mathcal{W} = \mathcal{H}_1 \cap \mathcal{V}.
\end{equation}
It is worth noting that $\mathcal{W}$ is a convex set, thanks to the convexity of $\mathcal{H}_1$ and $\mathcal{V}$. 

\section{Framework for GRD Function Modeling}
\label{sec:computational}
In order to find a suitable parametrization of the infinite-dimensional space $\mathcal{W}$, we make use of the relations $\mathcal{H}_1 = f_0 + \mathcal{H}_0$ and $\mathcal{W} = \mathcal{H}_1 \cap \mathcal{V}$.
We first conclude that $\forall h \in \mathcal{H}_0$, $h$ is square-integrable as $h$ is a continuous function defined over a compact set as shown by Eq.~\eqref{eq:grd_h0}.
Therefore, it is possible to equip the space $\mathcal{H}_0$ with an inner product
\begin{equation}
    \langle h,g\rangle \coloneqq \iint_{\Omega}h(x,y)g(x,y)dxdy, \forall h, g \in \mathcal{H}_0, \label{eq:inner_prod}
\end{equation}
and define an induced metric by
\begin{equation}
	d_2(h, g) \coloneqq \left[ \iint_{\Omega}|h(x,y) - g(x,y)|^2dxdy \right]^{\frac{1}{2}}, \forall h, g \in \mathcal{H}_0. \nonumber
\end{equation}
With the metric $d_2$ at hand, we may complete $\mathcal{H}_0$ by including the limits of all Cauchy sequences that belong to the functional subspace. 
It turns out that the completion of $\mathcal{H}_0$ is the space of all square-integrable functions defined on $\Omega$, which we denote by $\mathcal{L}_2(\Omega)$.
By definition, $\mathcal{L}_2(\Omega)$ is a Hilbert space with Eq.~\eqref{eq:inner_prod} being the inner product operation, and $\mathcal{H}_0$ is a dense subset of $\mathcal{L}_2(\Omega)$~\cite{kreyszig1978introductory}.

Then we are able to model $\mathcal{W}$ with countable parameters. 
It is known that $\mathcal{H}_0$ is separable, as polynomial functions form a dense countable subset of $\mathcal{H}_0$~\cite{kreyszig1978introductory}. 
Therefore, we conclude that there exists an orthonormal basis $\{h_n, n=1,2,3,\cdots\} \subset \mathcal{H}_0$ that spans $\mathcal{L}_2(\Omega)$:
\begin{equation}
    h \sim \sum_{n=1}^{\infty}c_nh_n,~~\forall h \in \mathcal{L}_2(\Omega) \label{eq:expansion}
\end{equation}
where $\sim$ denotes the equality relationship in the $d_2$ sense, and $c_n = \langle h,h_n \rangle \in \mathbb{R}$.
As a result, any GRD function $f \in \mathcal{W}$ can be expressed as a linear combination of $\{h_n\}$:
\begin{equation}
\label{eq:W_grd_basis}
\exists \{c_n\}, \text{such that } f = f_0 + \sum_{n=1}^{\infty}c_nh_n, \forall f \in \mathcal{W}.
\end{equation}
Eq.~\eqref{eq:W_grd_basis} not only parametrizes the theoretical space of GRD functions, but also provides a series of approximation models.
For example, it is straightforward to compute an $N$-th order approximation:
\begin{equation}\label{eq:general_model}
    \tilde{f} = f_0 + \sum_{n=1}^{N} c_n h_n,
\end{equation}
As $N$ becomes larger, the above model approximates the GRD functions better in $\mathcal{W}$.

The parametrization of GRD functions also provides a systematic way of estimating a GRD function from samples.
The $N$-th order model in Eq.~\eqref{eq:general_model} defines an $N$-dimensional approximation of $\mathcal{W}$:
\begin{equation}
    \tilde{\mathcal{W}}_N \coloneqq \left\{f\bigg{|} f = f_0 + \sum_{n=1}^{N}c_nh_n, f \in \mathcal{V} \right\}. \label{eq:tilde_wgrd}
\end{equation}
The approximation space $\tilde{\mathcal{W}}_N$ is a subset of $\mathcal{W}$ as $\{h_n\} \subset \mathcal{H}_0$, meaning that any element in $\tilde{\mathcal{W}}_N$ is a valid GRD function.
Therefore, estimating a GRD function corresponds to finding the optimal element in $\tilde{\mathcal{W}}_N$ that best fits given samples. Since $\tilde{\mathcal{W}}_N$ is a closed convex set, we formulate GRD function estimation as the projections-onto-convex-sets (POCS) problem.
Given a set of attribute-quality pairs $\{f(x_i, y_i) = z_i, i \in \mathcal{I}\}$, where $\mathcal{I}$ denotes the index set, we aim to solve
\begin{equation}
    \label{eq:proj}
    \begin{split}
        \argmin_{\{c_n\}} & ~~ \sum_{i \in \mathcal{I}} |z_i - f_0(x_i, y_i) - \sum_{n=1}^{N}c_nh_n(x_i, y_i)|^2 \\
        \text{s.t.} & ~~ f_0 + \sum_{n=1}^{N}c_nh_n \in \mathcal{V}.
    \end{split}
\end{equation}
We then plug the optimal coefficients $\{c_n^*\}$ into Eq.~\eqref{eq:general_model} to obtain the estimated GRD function.


\section{eGRD Model for GRD Functions}
The general framework proposed in Section~\ref{sec:computational} allows arbitrary orthonormal basis, which leads to different GRD methods. For example, a polynomial model can be obtained by setting $h_n$ in Eq.~\eqref{eq:general_model} to the 2-dimensional polynomial basis. 
Similarly, one can obtain a trigonometric approximation model with $h_n$ being the half-sine basis.
One drawback of fixed basis functions is that they are not adaptive to  real-world GRD functions, and thus may not capture the directions of large variations on the data manifold. Such models may need a great number of basis functions to achieve a satisfactory approximation accuracy, which in turn require a great number of training samples.
In this section, we seek a minimal set of eigen basis functions that can effectively represent the majority of real-world GRD functions. The resulting eigen GRD (eGRD) model for GRD functions is data efficient, meaning that it can be accurately estimated from sparse samples.

\subsection{Optimal Basis of Real-World GRD Functions}\label{sec:real}
\begin{figure*}
\centering
\includegraphics[width=0.8\textwidth]{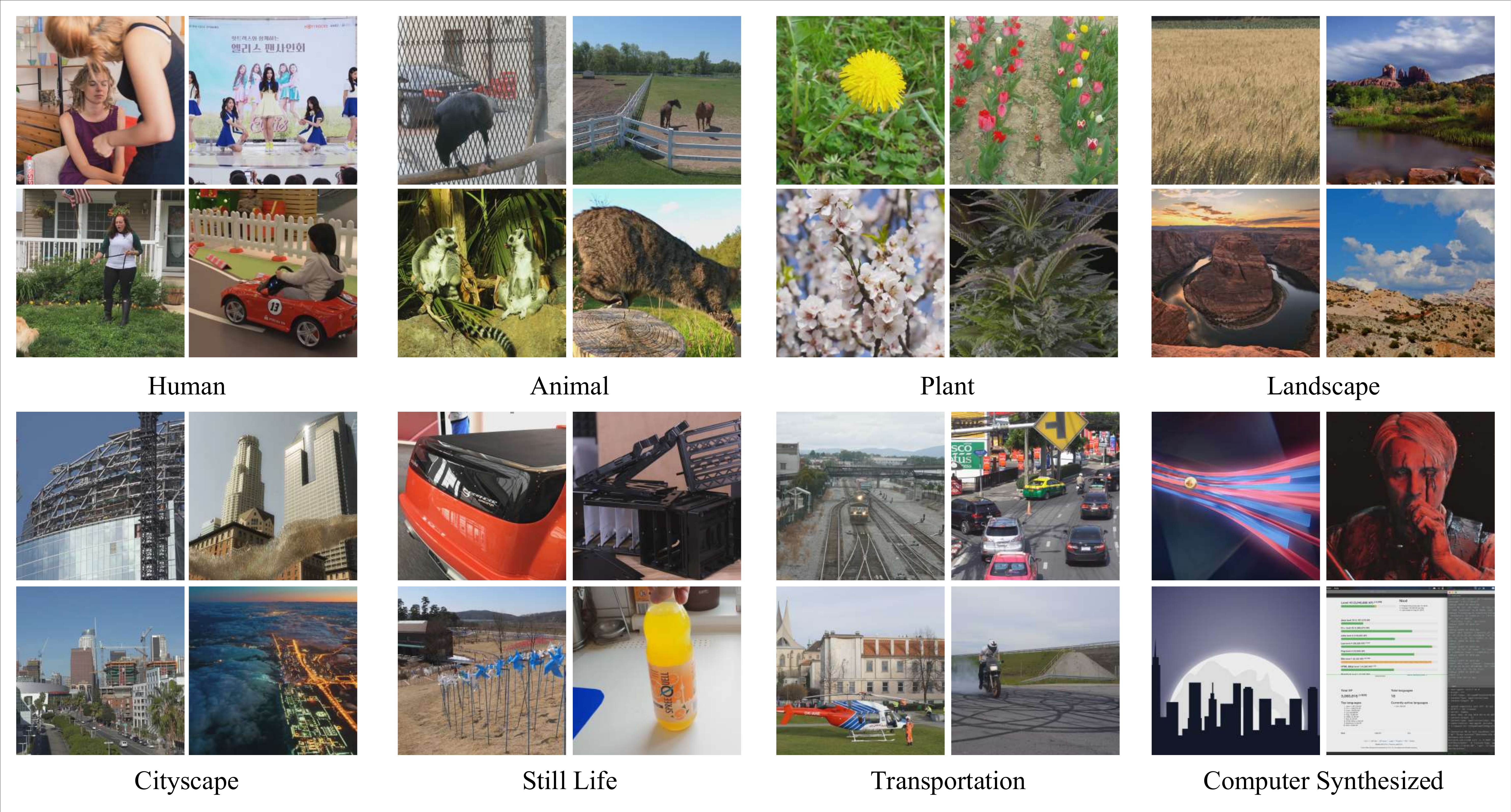}
\caption{Sample frames of source videos in the Waterloo GRD database. All images are cropped for neat presentation.}
\label{fig:sample_video}
\end{figure*}

\subsubsection{GRD Function Database}\label{subsec:database}
Although the GRD function is continuous in theory, we often work with a discrete version in practice. For example, a limited number of profiles are specified in video encoding, suggesting that only a finite number of samples on a GRD surface are practically achievable.
Here, we densely sample the bitrate-resolution space on a rectangular grid, and collect all the GRD function values (\textit{i.e.}, quality of corresponding representations) as a $K$-dimensional vector.
Hereafter, we treat  $\mathbf{f} \in \mathbb{R}^K$ as the ground-truth discretization of the GRD function $f$, with the mild assumption that $f$ is smooth enough to be recovered from its dense samples\footnote{In fact, when the GRD function is band-limited, it can be fully recovered with the Nyquist rate.}.

Following this idea, we construct a large-scale database of GRD functions, namely the Waterloo GRD database.
First, we collect $1,000$ pristine videos with Creative Commons licenses, spanning a great diversity of video content. 
To make sure that the selected videos are of pristine quality, 
we perform two rounds of screening to remove those videos with visible distortions.
We further reduce any possible artifacts by
downsampling the videos to the size of $1,920\times 1,080$, from which we trim $10$-second semantically coherent video clips. 
Eventually, we end up with $1,000$ high-quality $10$-second videos. 
Sample frames are shown in Fig.~\ref{fig:sample_video}, where we can see the richness of video content.

Each video in the database is distorted by the following sequential process:
\begin{itemize}
  \item Spatial downsampling: We downsample the source video using the bicubic filter to six spatial resolutions ($1920 \times 1080$, $1280 \times 720$, $720 \times 480$, $512 \times 384$, $384 \times 288$, $320 \times 240$) according to the list of Netflix certified devices~\cite{de2016complexity}. 
  Consequently, the lower and upper bounds of spatial resolution  are $y_{\min} = 400$ and $y_{\max} = 2203$, respectively.
  \item H.264/VP9 compression: We encode the downsampled sequences using two commonly used video encoders, \textit{i.e.}, H.264 and VP9, with two-pass encoding~\cite{grois2013performance,de2016complexity,kreuzberger2016comparative}. 
  The target bitrate ranges from 100 kbps to 9 Mbps with a step size of 100 kbps.
  Thus the lower and upper bounds of bitrate are $x_{\text{min}} = 100$ kbps and $x_{\text{max}} = 9000$ kbps, respectively. 
  The full encoding specification is detailed in Appendix~\ref{sec:encoding_cmd}.
\end{itemize}
In total, we obtain $540$ (hypothetical reference circuit~\cite{bt500subjective}) $\times$ $1,000$ (content) $\times$ $2$ (encoder) $= 1,080,000$ video representations (currently the largest in the VQA literature). 
Ideally, the response of a GRD function should be measured by subjective evaluation, because the human eye is the ultimate receiver in most visual applications. 
However,  subjective testing is expensive and time consuming. 
Here we opt to replace subjective assessment with objective VQA measurements. 
Specifically, we use SSIMplus~\cite{Rehman2015SSIMplus} to evaluate the quality of the $1,080,000$ video representations for the following reasons. 
First, SSIMplus is shown to outperform other state-of-the-art quality measures in terms of accuracy and speed~\cite{Rehman2015SSIMplus,Duanmu2017QoE}. 
Second, it is currently the only objective VQA model that offers meaningful cross-resolution and cross-device scoring.
Third, its precedent models SSIM~\cite{wang2004image} and MS-SSIM~\cite{wang2003multiscale} have been demonstrated to perform well in estimating GRD functions~\cite{chen2016subjective}, and have been widely used in industry practice. 
The outputs of SSIMplus are regarded as the ground-truth responses of GRD functions. The range of SSIMplus is from $0$ to $100$, with $100$ indicating perfect quality.
It is worth noting that our GRD modeling approach does not restrict itself to any specific VQA method. 

We post-process the raw data to obtain GRD functions on a regular grid. 
First, the lossless encoding bitrate may be lower than $9,000$ kbps when the complexity of the source video is low.
In such case, we pad the highest achievable quality at each resolution to the end of the GRD function along the bitrate dimension. 
Second, the rate-control of video encoders may be inaccurate, leading to discrepancies between the actual and the target encoding bitrates. 
To resolve this, we resample the rate-distortion curves at each resolution uniformly with a step-size of $100$ kbps using 1D piecewise cubic Hermite interpolation to preserve monotonicity.
In the end, we obtain $2,000$ GRD functions from $1,000$ source videos and $2$ video encoders.
The large variations of the GRD functions due to content diversity are shown in Fig.~\ref{fig:sample_grd}.

\subsubsection{Eigen Basis for Real-World GRD Functions}
\begin{figure}
	\centering
	\includegraphics[width=0.45\textwidth]{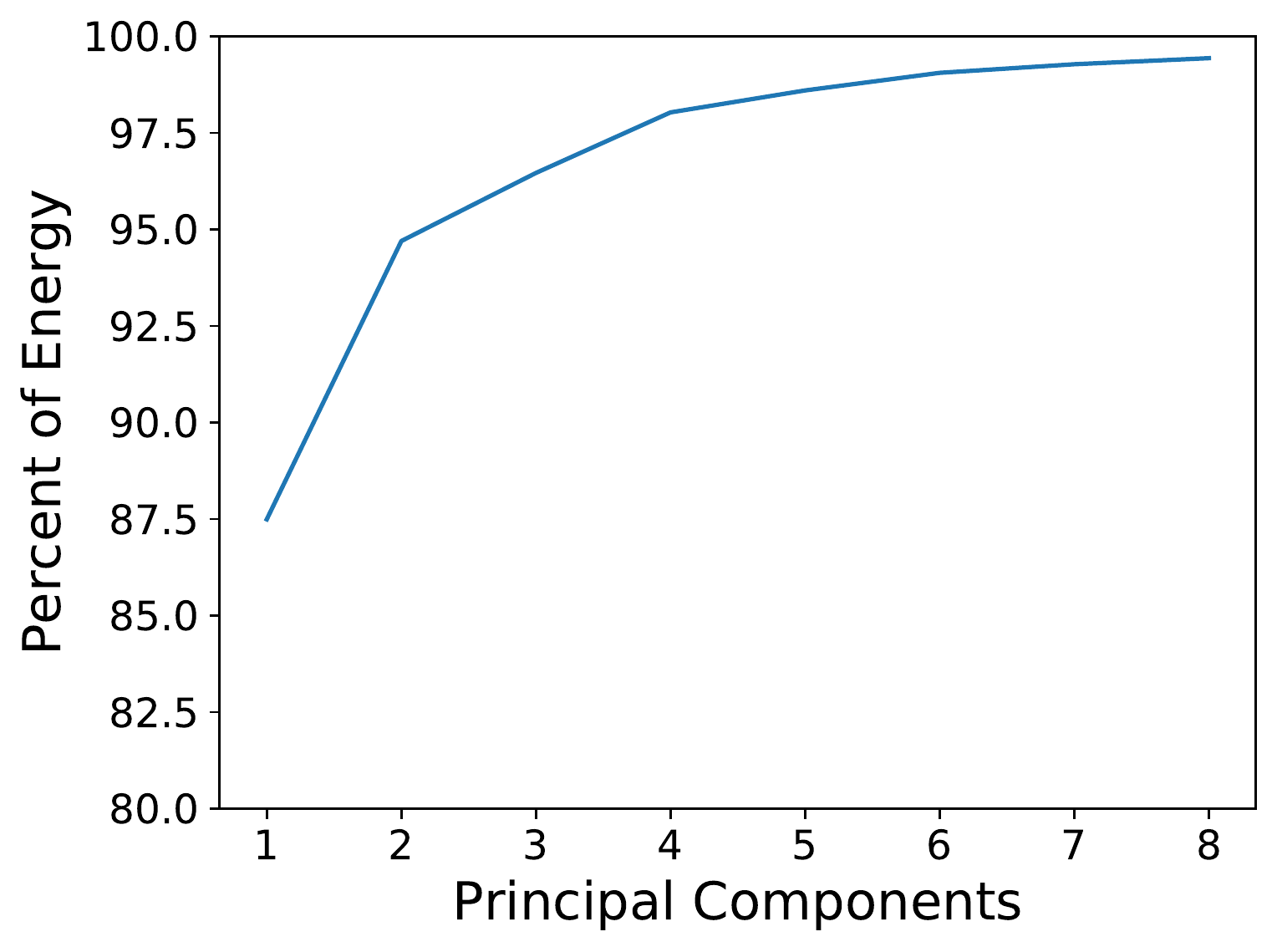}
 	\caption{The percentage of the energy explained by the span of the first $N$ principal components. 
}
	\label{fig:eigen_engergy}
\end{figure}
\begin{figure*}
	\centering
	\subfloat[Mean GRD surface]{\includegraphics[width=0.24\textwidth]{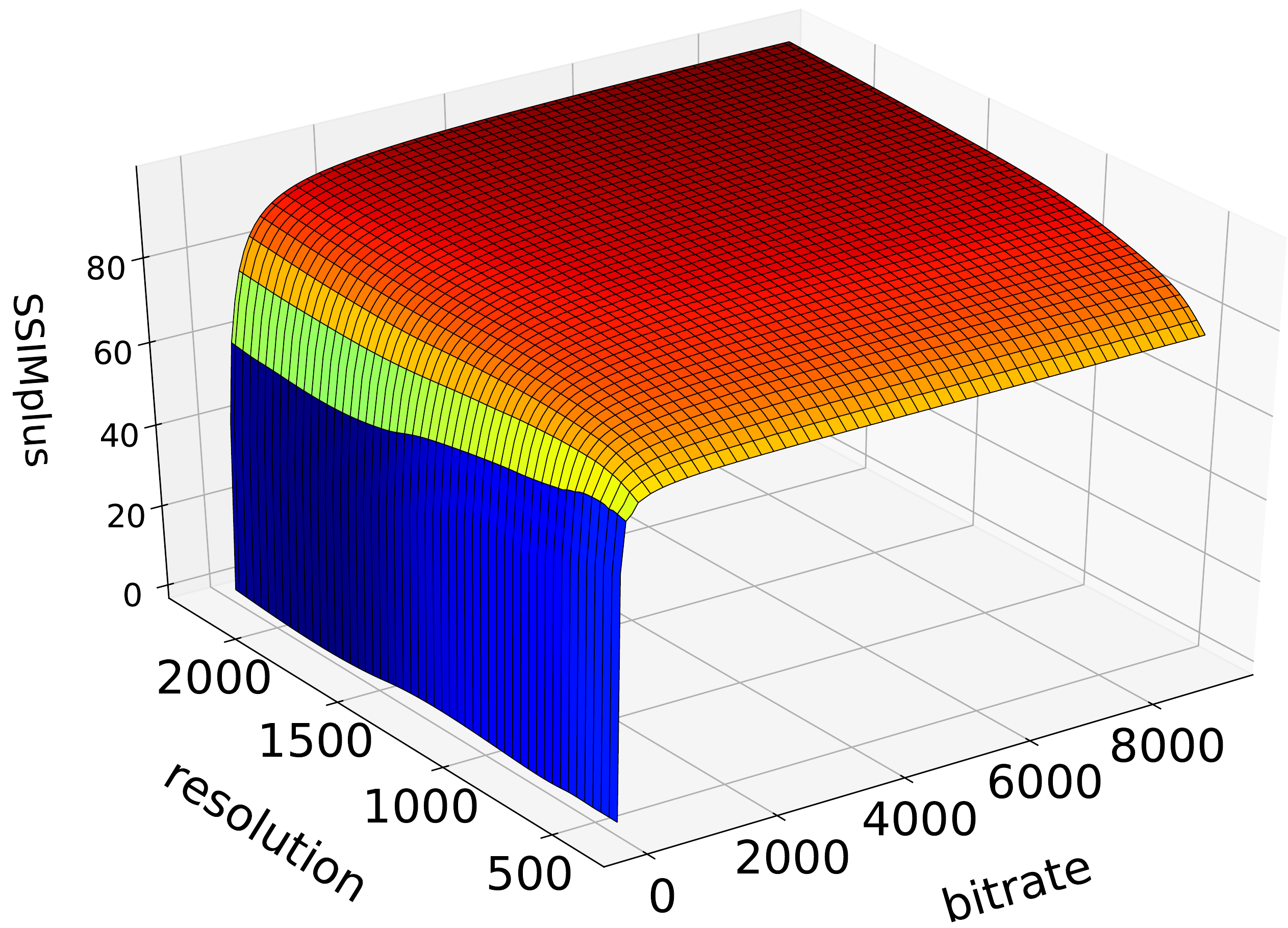}}
	\subfloat[First principal component]{\includegraphics[width=0.24\textwidth]{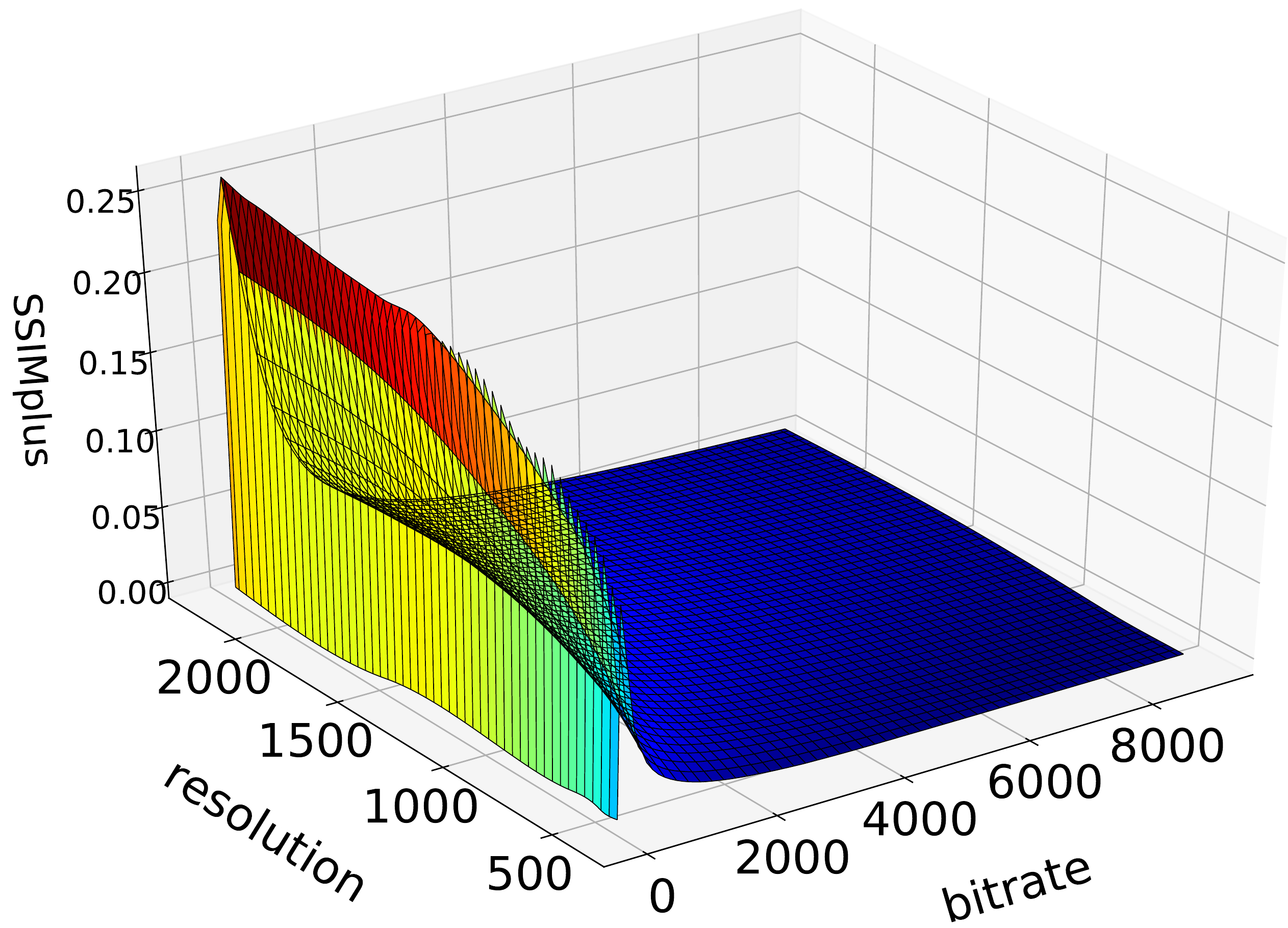}}
	\subfloat[Second principal component]{\includegraphics[width=0.24\textwidth]{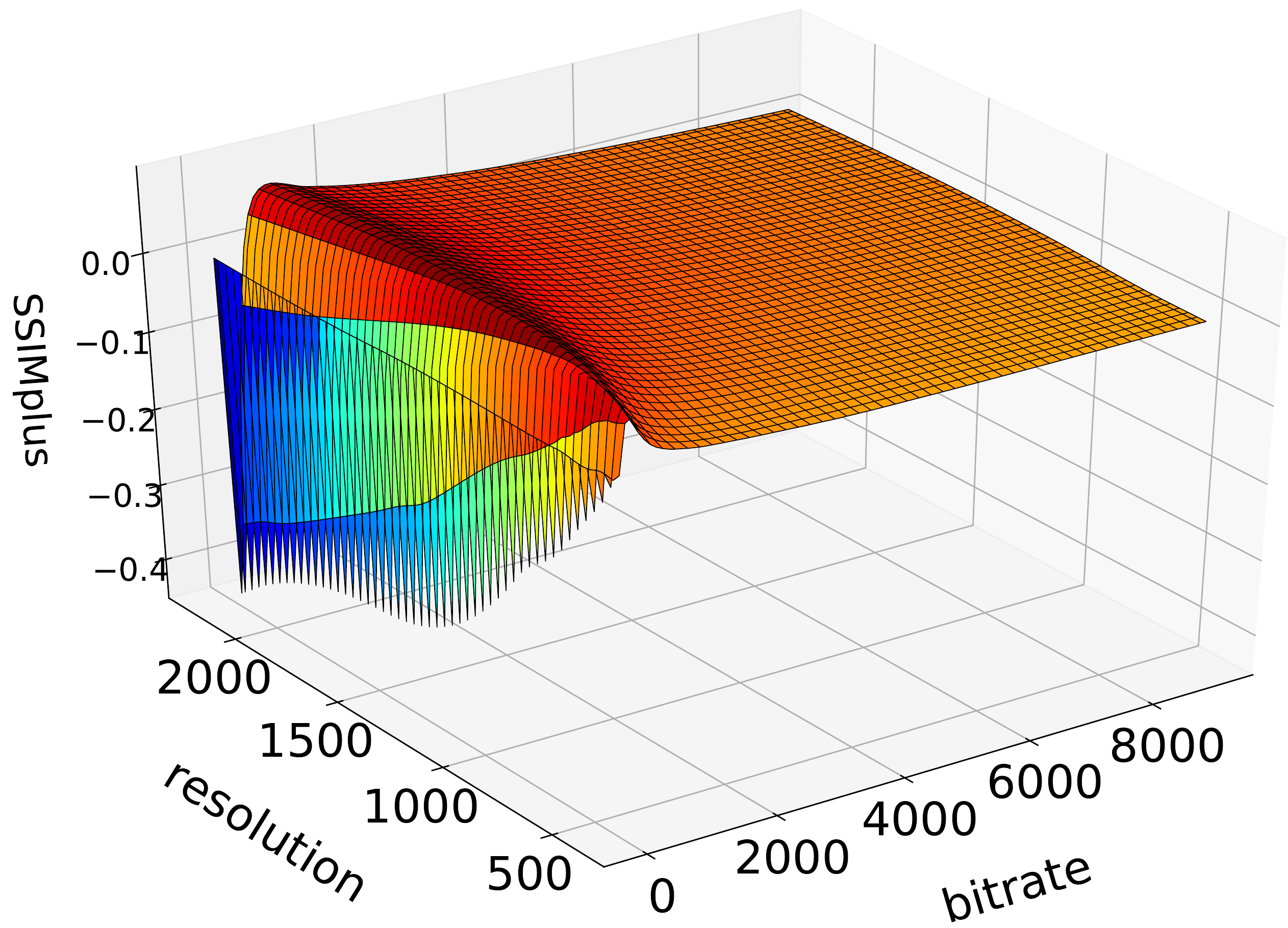}}
	\subfloat[Third principal component]{\includegraphics[width=0.24\textwidth]{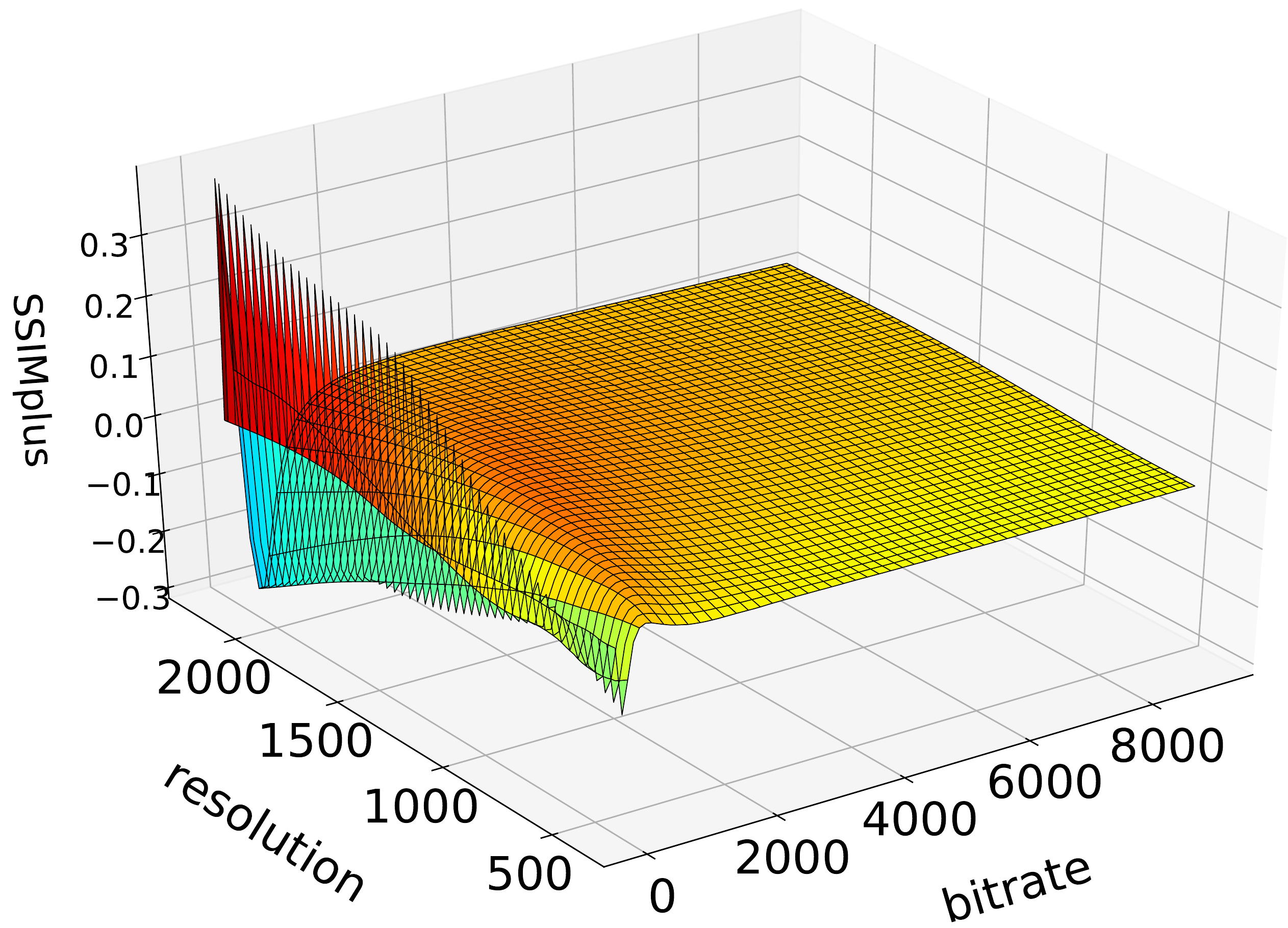}}\\
	\subfloat[Fourth principal component]{\includegraphics[width=0.24\textwidth]{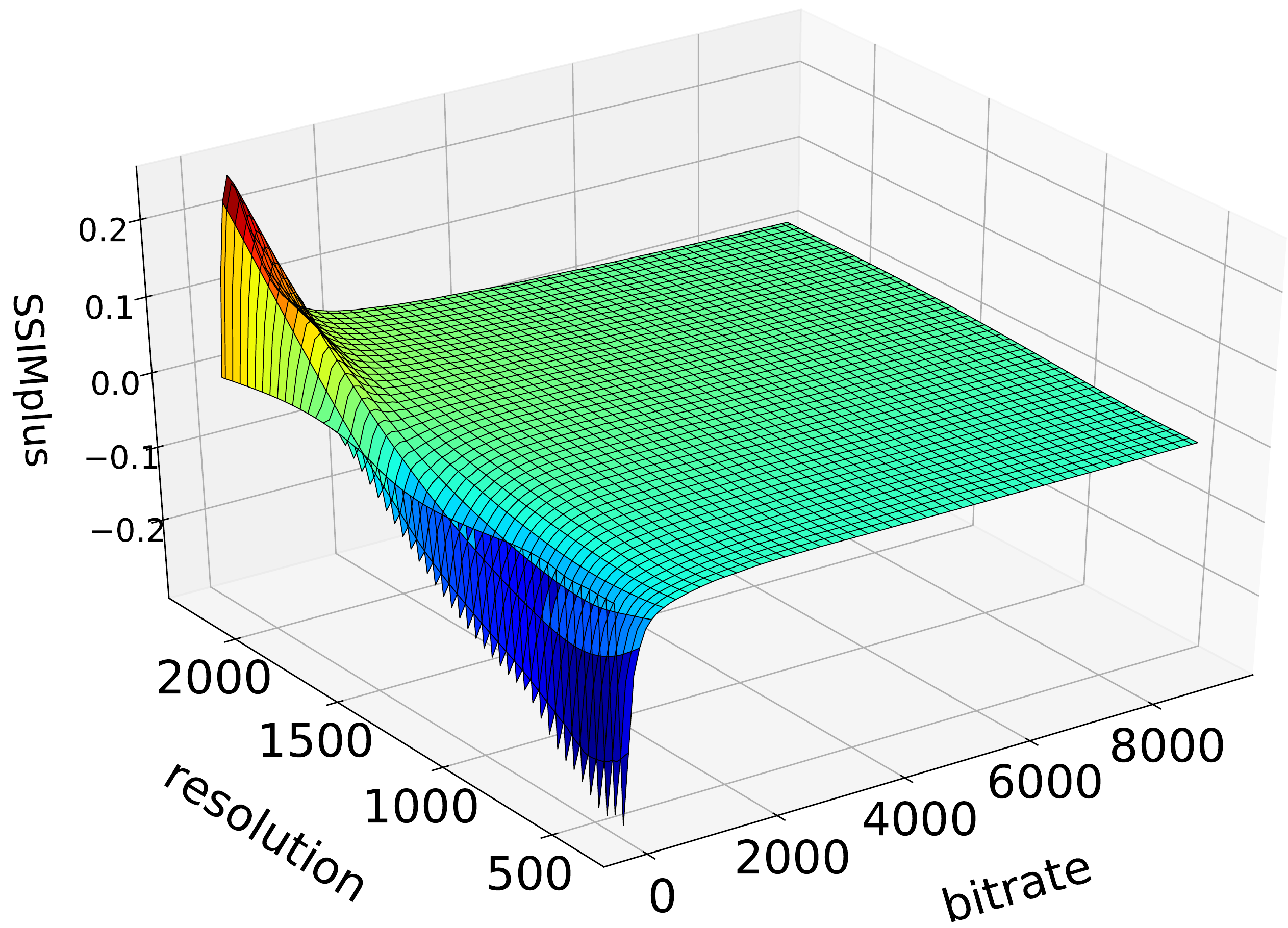}}
	\subfloat[Fifth principal component]{\includegraphics[width=0.24\textwidth]{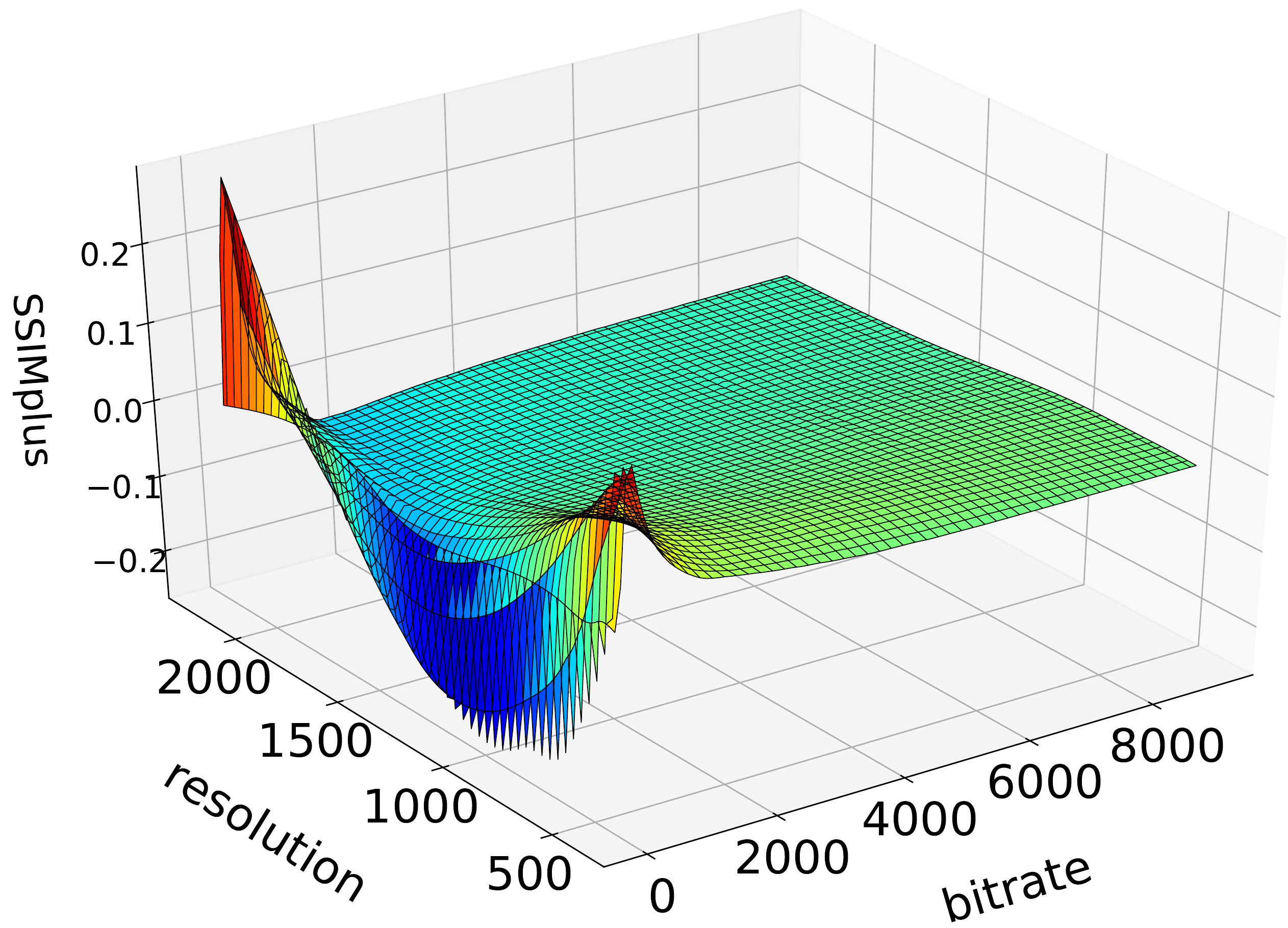}}
	\subfloat[Sixth principal component]{\includegraphics[width=0.24\textwidth]{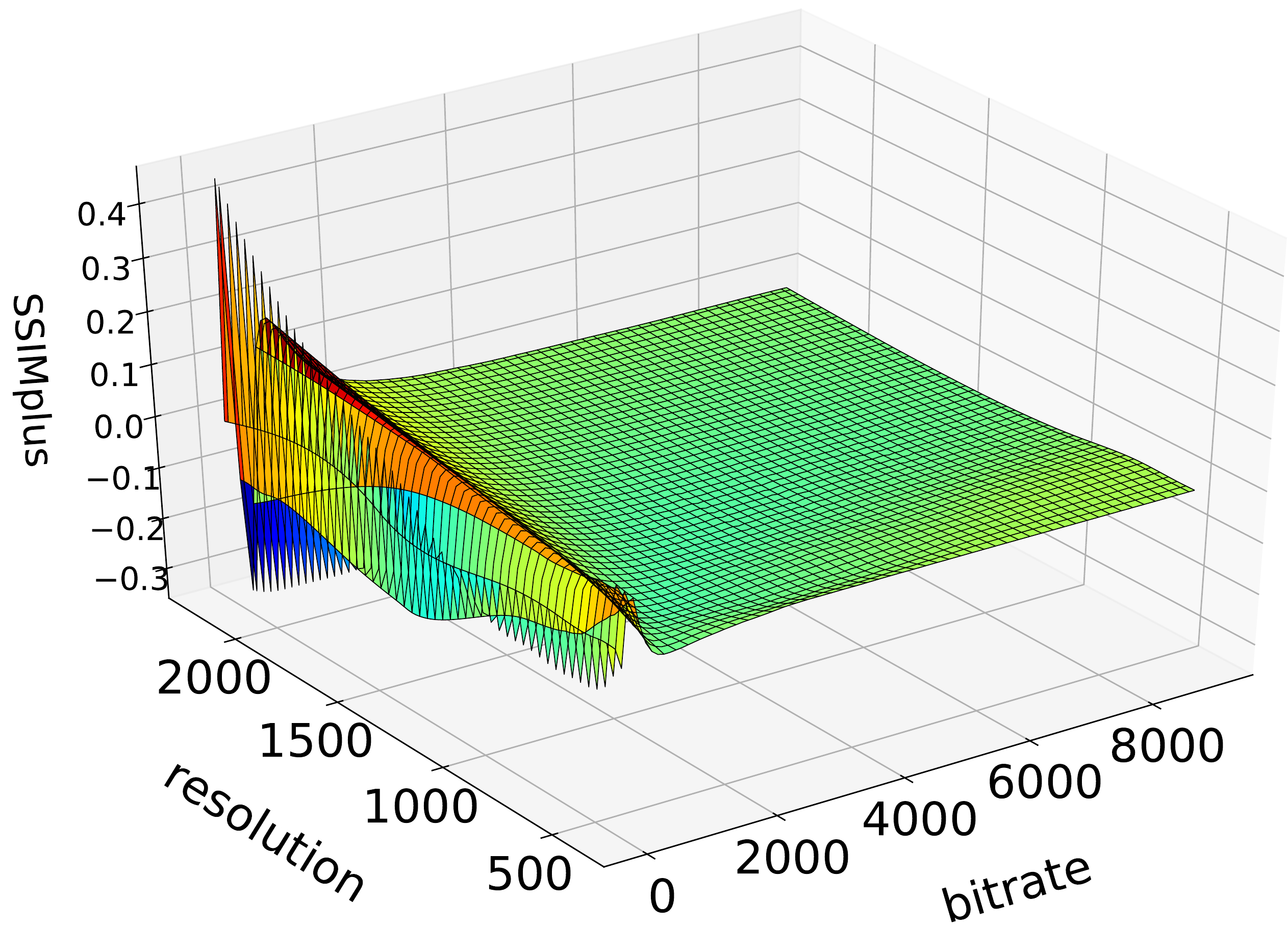}}
	\subfloat[Seventh principal component]{\includegraphics[width=0.24\textwidth]{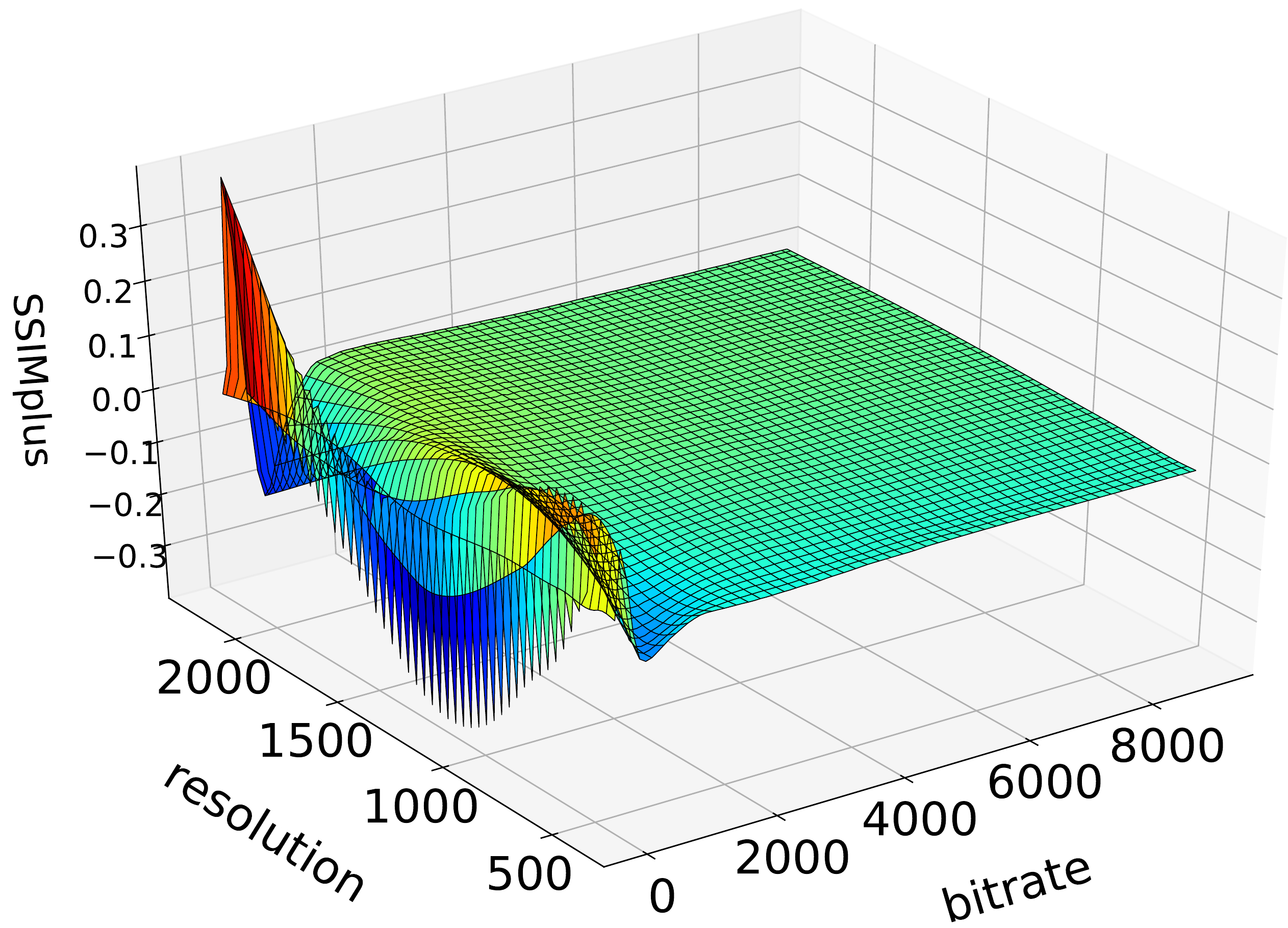}}
	\caption{The mean and  the first seven principal components of real-world GRD functions.}
	\label{fig:eigen_surfaces}
\end{figure*}
Recall that we aim to discover a set of basis that best approximate the real-world GRD functions.
Consider the $m$-th real-world GRD function, denoted by $\mathbf{f}_m$, in the Waterloo GRD database, whose
 best approximation using the $N$-th order model in Eq.~\eqref{eq:general_model} is achieved by
\begin{equation}
	\tilde{\mathbf{f}}_m \coloneqq \mathbf{f}_0 + \sum_{n=1}^N \langle \mathbf{f}_m - \mathbf{f}_0, \mathbf{h}_n\rangle \mathbf{h}_n, \nonumber
\end{equation}
with an approximation error given by
\begin{equation}
    \mathcal{E}[\mathbf{f}_m] \coloneqq \left|\mathbf{f}_m - \left( \mathbf{f}_0 + \sum_{n=1}^N \langle \mathbf{f}_m - \mathbf{f}_0, \mathbf{h}_n\rangle \mathbf{h}_n \right) \right|_2, \label{eq:approx_error}
\end{equation}
where $|\cdot|_2$ indicates the Euclidean norm of a vector, and $\mathbf{h}_n$ denotes the discrete version of the basis function $h_n$.
Given $M$ empirical GRD functions in the Waterloo GRD database, the optimal orthonormal basis is thus obtained by minimizing the average approximation error:
\begin{equation}
\small
\begin{split}
\argmin_{\mathbf{f}_0, \{\mathbf{h}_n\}} &~~ \frac{1}{M}\sum_{m=1}^{M}\left|\mathbf{f}_m - \mathbf{f}_0- \sum_{n=1}^{N}\langle \mathbf{f}_m - \mathbf{f}_0, \mathbf{h}_n\rangle \mathbf{h}_n\right|_2^2,\\        
\text{s.t.} &~~ |\mathbf{h}_n|_2^2 = 1,~~n=1,\cdots,N,\\
&~~ \langle\mathbf{h}_n,\mathbf{h}_{n'}\rangle = 0,~~n,n' \in \{1,\cdots,N\}, n \neq n'.
\end{split}\label{eq:pca}
\end{equation}
For the case of $N=0$, it is trivial to show that the optimal $\mathbf{f}_0^*$ equals the mean of the $M$ GRD functions, which is valid due to the convexity of $\mathcal{W}$.
When $N \geq 1$, Problem~\eqref{eq:pca} is essentially principal component analysis (PCA), meaning that the $n$-th optimal component $\mathbf{h}_n^*$ is the eigenvector associated with the $n$-th largest eigenvalue of the empirical covariance matrix of $\mathbf{f}_m$.
The optimal $N$-dimensional approximation of $\mathcal{W}$ is also achieved by the span of the first $N$ eigenvectors plus $\mathbf{f}_0^*$.

Fig.~\ref{fig:eigen_engergy} shows that the cumulative energy explained  increases rapidly with the number of principal components. In fact, eight components explain more than $99.5\%$ of the energy.
This suggests that most real-world GRD functions lie in a low-dimensional subspace, and that the resulting eGRD models with only a few parameters  should work  well. In order to gain an impression about the shapes of the eigen GRD functions, we visualize the mean GRD surface $\mathbf{f}_0^*$ and the first seven empirical principal components $\mathbf{h}_1^*$ to $\mathbf{h}_7^*$ in Fig.~\ref{fig:eigen_surfaces}, from which we have two observations.
First, among the seven principal components, the first one is the smoothest, while the second to the seventh are increasingly oscillatory.
This finding implies that the perceptual quality of a video representation is positively correlated with its neighboring representations in general.
Second, all the principal components exhibit the greatest magnitudes in regions with low bitrate and high resolution, indicating their complicated combined effects on perceptual quality. 

\subsection{eGRD Model Estimation from Sparse Samples}
\label{sec:recon}
By inserting the learned mean $\mathbf{f}_0^*$ and the principal components $\{\mathbf{h}_n^*\}$ into the POCS problem, the parameters of our eGRD model can be efficiently estimated  from sparse samples.
Specifically, to make Problem \eqref{eq:proj} practically solvable, we approximate its constraints as a set of linear inequalities.
 We first rewrite Eq.~\eqref{eq:general_model} in matrix form
\begin{equation}
\label{eq:egrd_mat}
    \tilde{\mathbf{f}} = \mathbf{f}_0 + H_N^*\mathbf{c},
\end{equation}
where $H_N^* \coloneqq [\mathbf{h}_1^*, \mathbf{h}_2^*, \cdots, \mathbf{h}_N^*]$ and $\mathbf{c} \coloneqq [c_1,c_2, \cdots, c_N]^T$.
Denote the matrix of  the first order difference along the $x$-axis by $D_x$, and the matrix of the first order difference along the $y$-axis only when $x=x_{\max}$ by $D_y$, respectively.
The discrete form of $\mathcal{V}$ can be expressed by
\begin{equation}
\label{eq:discrete_V}
\left[ {\begin{array}{*{20}{c}}
	{{D_x}}\\
	{{D_y}}
	\end{array}} \right]\tilde{\mathbf{f}} \geq 0.
\end{equation}
By substituting~\eqref{eq:egrd_mat} into~\eqref{eq:discrete_V}, we obtain
\begin{equation}
\label{eq:disc_V_in_c}
- \left[ {\begin{array}{*{20}{c}}
	{{D_x}}\\
	{{D_y}}
	\end{array}} \right]{\mathbf{H}_N^*}\mathbf{c} \leq \left[ {\begin{array}{*{20}{c}}
	{{D_x}}\\
	{{D_y}}
	\end{array}} \right]{\mathbf{f}_0},
\end{equation}
which imposes linear constraints on the coefficients $\mathbf{c}$.
As a result, finding optimal $\mathbf{c}^*$ turns into a quadratic programming problem, which can be solved by convex optimization tools, such as OSQP~\cite{stellato2017osqp}.
Finally, by substituting $\mathbf{c}^*$ into Eq.~\eqref{eq:general_model}, we obtain the best eGRD model that fits known samples with least squared errors.

\section{Experiments}\label{sec:performance}
In this section, we first quantitatively evaluate the approximation capability of the proposed eigen basis on the Waterloo GRD database. 
Then, we compare the performance of the eGRD model with existing methods on reconstructing GRD functions from sparse samples.
Furthermore, extensive experiments are conducted to show the robustness of the eGRD method in various practical scenarios.
Finally, we rely on another VQA model~\cite{li2016VMAF} to demonstrate the  generality of the eGRD method.

\subsection{Approximation Capability of Basis}
\begin{table}[t!]
	\centering
	\caption{Mean and worst performance of eGRD on the training set with different numbers of basis functions}
	\label{tab:egrd_training}
	\begin{tabular}{c|cc|cc}
		\toprule
		\multirow{2}{*}{$N$} & \multicolumn{2}{c|}{RMSE} & \multicolumn{2}{c}{$l^\infty$ error}\\ \cline{2-5}
		& Mean & Worst & Mean  & Worst \\
		\hline
		0 & 3.88 & 20.18 & 29.66 & 74.77 \\
		1 & 1.80 & 14.07 & 17.43 & 50.54 \\
		2 & 1.08 & 7.71 & 9.64 & 42.81 \\
		3 & 0.91 & 4.85 & 7.92 & 36.60 \\
		4 & 0.77 & 4.82 & 6.69 & 36.42 \\
		5 & 0.61 & 4.08 & 5.21 & 24.18 \\
		6 & 0.45 & 3.69 & 4.06 & 25.36 \\
		7 & 0.41 & 2.48 & 3.41 & 24.02 \\
		8 & 0.37 & 2.23 & 2.88 & 14.65 \\
		\bottomrule
	\end{tabular}
\end{table}
\begin{table}[t!]
	\centering
	\caption{RMSE of GRD models with different basis functions on the test set. Best results for mean and worst performance are highlighted in italics and boldface, respectively}\label{tab:basis_rmse}
		\begin{tabular}{c|cc|cc|cc}
			\toprule
			\multirow{2}{*}{$N$}   & \multicolumn{2}{c|}{Polynomial} & \multicolumn{2}{c|}{Trigonometric} & \multicolumn{2}{c}{Eigen} \\ \cline{2-7}                                  
			~ & Mean & Worst & Mean & Worst & Mean & Worst \\ \hline
			0 & \textit{3.86} & \textbf{18.06} & \textit{3.86} & \textbf{18.06} & \textit{3.86} & \textbf{18.06} \\
			1 & 3.81 & 17.90 & 3.83 & 17.98 & \textit{1.79} & \textbf{11.67} \\
			2 & 3.72 & 15.91 & 3.75 & 17.02 & \textit{1.09} & \textbf{5.29} \\
			3 & 3.68 & 14.88 & 3.68 & 16.52 & \textit{0.92} & \textbf{3.83} \\
			4 & 3.65 & 14.86 & 3.67 & 16.03 & \textit{0.77} & \textbf{3.29} \\
			6 & 3.19 & 10.26 & 3.57 & 15.26 & \textit{0.46} & \textbf{2.54} \\
			8 & 2.86 & 8.97 & 3.44 & 14.41 & \textit{0.38} & \textbf{2.14} \\
			10 & 2.55 & 8.34 & 3.37 & 13.89 & \textit{0.32} & \textbf{1.65} \\
			15 & 1.92 & 7.70 & 3.17 & 12.51 & \textit{0.26} & \textbf{1.38} \\
			20 & 1.83 & 6.10 & 3.02 & 11.54 & \textit{0.20} & \textbf{1.09} \\
			\bottomrule
	\end{tabular}
\end{table}
\begin{table}[t!]
	\centering
	\caption{$l^\infty$ error of GRD models with different basis functions on the test set}\label{tab:basis_linf}
		\begin{tabular}{c|cc|cc|cc}
			\toprule
			\multirow{2}{*}{ $N$}   & \multicolumn{2}{c|}{Polynomial} & \multicolumn{2}{c|}{Trigonometric} & \multicolumn{2}{c}{Eigen} \\ \cline{2-7}                                  
			~ & Mean & Worst & Mean & Worst & Mean & Worst \\ \hline
			0 & \textit{29.67} & \textbf{66.97} & \textit{29.67} & \textbf{66.97} & \textit{29.67} & \textbf{66.97} \\
			1 & 29.63 & 66.87 & 29.65 & 66.92 & \textit{17.54} & \textbf{45.29} \\
			2 & 29.46 & 63.80 & 29.58 & 66.23 & \textit{9.67} & \textbf{37.42} \\
			3 & 29.30 & 59.71 & 29.47 & 65.51 & \textit{8.01} & \textbf{34.48} \\
			4 & 29.13 & 59.69 & 29.43 & 64.65 & \textit{6.76} & \textbf{33.68} \\
			6 & 27.32 & 54.51 & 29.18 & 62.82 & \textit{4.12} & \textbf{24.40} \\
			8 & 25.62 & 53.21 & 28.79 & 60.22 & \textit{2.94} & \textbf{14.57} \\
			10 & 23.55 & 50.86 & 28.56 & 57.96 & \textit{2.37} & \textbf{12.34} \\
			15 & 17.12 & 40.88 & 27.72 & 55.12 & \textit{1.69} & \textbf{9.61} \\
			20 & 16.71 & 40.72 & 27.07 & 54.36 & \textit{1.14} & \textbf{7.44} \\
			\bottomrule
	\end{tabular}
\end{table}

As discussed in Section \ref{sec:computational}, we may change the proposed eigen basis to the polynomial or the trigonometric basis in Eq.~\eqref{eq:general_model}, resulting in two alternative GRD models - polynomial GRD (pGRD) and trigonometric GRD (tGRD).
All the models can fit increasingly complex GRD functions at the cost of more basis functions and coefficients.
What distinguishes them is the rate at which the approximation error diminishes as the number of basis vectors increases.
We compute four kinds of approximation errors on the Waterloo GRD database. Specifically,
for each GRD surface, we calculate the root-mean-square error (RMSE) and the $l^\infty$ error between the reconstructed and the ground-truth functions.
For a set of GRD functions, the average and the largest RMSE or $l^\infty$ errors are reported as the mean and the worst case performance of a GRD model. In order to train the principle components of the proposed eGRD model, we randomly split the database into a training set of $1,600$ GRD functions from $800$ source videos and a test set of the remaining $400$ GRD functions.
There is no content overlap between the training and test sets.
The random splitting is repeated $50$ times, and the median performance is reported. Besides, we use all samples of a GRD function to fit the model coefficients in this experiment.

First, we quantitatively evaluate how well the training data are represented by the learned eigenvectors. Table~\ref{tab:egrd_training} shows the reconstruction accuracy for $N=0,1,\ldots,8$, where $N=0$ means that all the GRD functions are approximated by the mean $f_0$.
As seen in the table, the trend is clear that the approximation capability improves as the number of basis vectors increases.
In particular, the training data can be precisely described by an eight-parameter eGRD model, with the RMSE  reduced to $0.37$.
According to previous studies~\cite{ma2017dipiq,gao2015learning}, such small quality differences are often regarded as indistinguishable to the human eye~\cite{bt500subjective}.
Moreover, the learned eigen basis can represent most eccentric GRD functions as indicated by an $l^\infty$ error as  small as $2.23$.
Another interesting finding is that even three principal components can achieve an average RMSE less than $1$, suggesting that the real-world GRD function space is of rather low dimensionality.

To emphasize the importance of basis selection, we compare the eigen basis with the polynomial and trigonometric basis by evaluating the approximation error of eGRD, pGRD and tGRD on the test set.
From Tables~\ref{tab:basis_rmse} and~\ref{tab:basis_linf}, we find that the eigen basis significantly outperforms the two alternatives, especially when the number of basis vectors is small.
In fact, the approximation capability of $20$ polynomial or trigonometric vectors is beaten by that of two eigenvectors with a clear margin.
This suggests that the eigen basis is more representative than general fixed basis to describe  variations of GRD functions.
In addition, increasing the number of eigenvectors improves the worst-case performance significantly, while adding more polynomial or trigonometric vectors achieves much less improvement.
\begin{table*}[!t]
	\centering
	\caption{RMSE of GRD models with different sample numbers} \label{tab:grd_recon_rmse}
	\begin{tabular}{c|cc|cc|cc|cc|cc}
		\toprule
		\multirow{2}{*}{$S$}   & \multicolumn{2}{c|}{Reciprocal~\cite{toni2015optimal}} & \multicolumn{2}{c|}{Logarithmic~\cite{chen2016subjective}} & \multicolumn{2}{c|}{pGRD} & \multicolumn{2}{c|}{tGRD} & \multicolumn{2}{c}{eGRD} \\ \cline{2-11}                                  
		~ & Mean & Worst & Mean & Worst & Average & Worst & Average & Worst & Average & Worst\\ \hline
		8 & N.A. & N.A. & N.A. & N.A. & 3.32 & 9.77 & 4.90 & 11.69 & \textit{0.71} & \textbf{3.04} \\
		10 & N.A. & N.A. & N.A. & N.A. & 3.28 & 9.21 & 4.58 & 11.74 & \textit{0.64} & \textbf{2.71} \\
		20 & N.A. & N.A. & 11.75 & 26.99 & 3.05 & 9.17 & 4.14 & 11.79 & \textit{0.50} & \textbf{2.58} \\
		30 & 13.57 & 38.35 & 9.13 & 19.37 & 3.04 & 9.13 & 4.05 & 11.81 & \textit{0.48} & \textbf{2.53} \\
		40 & 11.47 & 32.06 & 6.84 & 13.38 & 2.96 & 9.05 & 4.01 & 11.76 & \textit{0.46} & \textbf{2.48} \\
		50 & 9.14 & 33.02 & 5.70 & 12.07 & 2.93 & 9.04 & 3.94 & 11.77 & \textit{0.45} & \textbf{2.46} \\
		\bottomrule
	\end{tabular}
\end{table*}

\begin{table*}
	\centering
	\caption{$l^\infty$ error of GRD models with different sample numbers} \label{tab:grd_recon_linf}
	\begin{tabular}{c|cc|cc|cc|cc|cc}
		\toprule
		\multirow{2}{*}{$S$}   & \multicolumn{2}{c|}{Reciprocal~\cite{toni2015optimal}} & \multicolumn{2}{c|}{Logarithmic~\cite{chen2016subjective}} & \multicolumn{2}{c|}{pGRD} & \multicolumn{2}{c|}{tGRD} & \multicolumn{2}{c}{eGRD} \\ \cline{2-11}                                  
		~ & Mean & Worst & Mean & Worst & Mean & Worst & Mean & Worst & Mean & Worst\\ \hline
		8 & N.A. & N.A. & N.A. & N.A. & 24.46 & 52.95 & 28.46 & 59.92 & \textit{5.64} & \textbf{29.51} \\
		10 & N.A. & N.A. & N.A. & N.A. & 24.47 & 52.94 & 28.43 & 59.93 & \textit{3.21} & \textbf{18.00} \\
		20 & N.A. & N.A. & 29.64 & 61.13 & 24.58 & 53.04 & 28.51 & 59.77 & \textit{2.62} & \textbf{15.46} \\
		30 & 33.18 & 61.63 & 22.46 & 43.52 & 24.68 & 53.06 & 28.54 & 59.98 & \textit{2.47} & \textbf{12.91} \\
		40 & 36.53 & 67.03 & 21.02 & 42.20 & 24.92 & 53.05 & 28.57 & 59.97 & \textit{2.47} & \textbf{13.89} \\
		50 & 31.56 & 65.77 & 21.12 & 42.50 & 24.99 & 53.05 & 28.60 & 60.02 & \textit{2.50} & \textbf{13.89} \\
		\bottomrule
	\end{tabular}
\end{table*}

\subsection{GRD Reconstruction from Sparse Samples}\label{subsec:grd_recon}
We test five GRD models including reciprocal regression~\cite{toni2015optimal}, logarithmic regression~\cite{chen2016subjective},  pGRD,  tGRD, and eGRD on the Waterloo GRD database.
The first two methods are designed only for 1D RD curve estimation, but we extend them for 2D GRD surface reconstruction by performing RD curve regression at each resolution.
For the latter three models, we set the basis number to eight.
To sample a GRD function, we adopt an information-theoretic sampling method~\cite{duanmu2019modeling}, which generates a fixed sample sequence that minimizes the uncertainty of the function (see Appendix~\ref{sec:uncertainty} for more details).
The convergence rate of each method is reflected by the reconstruction errors with different sample numbers $S$, increasing gradually from $8$ to $50$.
Similar to the previous experiment, $80\%$ of the GRD functions are randomly selected as the training set for estimating the eigen basis and the sampling order.
The remaining $20\%$ samples constitute the test set.
The random splitting is repeated $50$ times, and the median results are reported.

Tables~\ref{tab:grd_recon_rmse} and \ref{tab:grd_recon_linf} summarize the results, from which  we have several key observations.
First, the proposed eGRD method significantly outperforms the reciprocal and the logarithmic regression methods in both accuracy and convergence rate.
This may be because 1) the competing methods presume fixed functional forms, which are poorly matched with real-world GRD functions, and 2) they treat a GRD surface as an aggregation of many 1D RD curves, missing the opportunity to exploit the dependency among different resolutions.
Second, eGRD delivers the best performance among its variants, while  pGRD  performs slightly better than tGRD.
This is consistent with the approximation capability of their respective basis  in the previous experiment.
Third, the performance of pGRD, tGRD and eGRD does not improve much as the sample number increases, implying that the primary influential factor might be the underlying basis.
Fourth,  eGRD can precisely recover a GRD surface with merely eight samples, based on which the reciprocal and  logarithmic models fail to  initialize the fitting process.

\subsection{eGRD with Varying Number of Basis Functions}\label{subsec:exp_egrd_variable}
\begin{table}[t!]
	\centering
	\caption{Mean and worst performance of eGRD when the number of basis vectors is equal to the number of samples}
	\label{tab:egrd_variable}
	\begin{tabular}{c|cc|cc}
		\toprule
		\multirow{2}{*}{$N$/$S$}& \multicolumn{2}{c|}{RMSE} & \multicolumn{2}{c}{$l^\infty$ error}\\ \cline{2-5}
		  & Mean & Worst & Mean  & Worst \\
		\hline
		1 & 1.83 & 11.66 & 17.80 & 53.38 \\
		3 & 1.20 & 4.59 & 8.73 & 32.63 \\
		5 & 0.87 & 4.36 & 6.65 & 25.48 \\
		8 & 0.71 & 3.04 & 5.64 & 29.51 \\
		30 & 0.40 & 1.85 & 2.50 & 12.98 \\
		50 & 0.22 & 1.10 & 1.03 & 7.15 \\
		\bottomrule
	\end{tabular}
\end{table}

We fix the number of basis functions of eGRD to eight in the previous experiment.  
However, under this setting, eGRD may neither work well with fewer samples nor benefit from more probes on the GRD function.
Here, we evaluate the performance of eGRD when the number of basis vectors is  equal to the number of samples. 
From Table~\ref{tab:egrd_variable}, we find that with varying number of basis vectors, eGRD can reconstruct a GRD surface even with a single sample. 
Moreover, we generally observe significant performance gains when more samples and basis functions are available. 
Throughout the rest of the paper, we use eGRD with varying basis described here for performance evaluation.

\subsection{Importance of Monotonicity Constraints}
\label{subsec:exp_monotonicity}
\begin{table}[t!]
	\centering
	\caption{Mean and worst performance of eGRD without monotonicity constraints}
	\label{tab:egrd_no_mono}
	\begin{tabular}{c|cc|cc}
		\toprule
		\multirow{2}{*}{$N$/$S$} & \multicolumn{2}{c|}{RMSE} & \multicolumn{2}{c}{$l^\infty$ error}\\ \cline{2-5}
		  & Mean & Worst & Mean  & Worst \\
		\hline
		1 & 1.45 & 8.95 & 15.13 & 53.41 \\
		3 & 0.87 & 4.59 & 7.79 & 31.72 \\
		5 & 0.73 & 4.54 & 6.10 & 27.06 \\
		8 & 0.76 & 6.39 & 7.34 & 62.40 \\
		30 & 4.52 & 27.88 & 26.78 & 151.15 \\
		50 & 3.15 & 19.11 & 15.03 & 79.02 \\
		\bottomrule
	\end{tabular}
\end{table}
\begin{figure}
	\centering
	\subfloat[]{\includegraphics[width=0.24\textwidth]{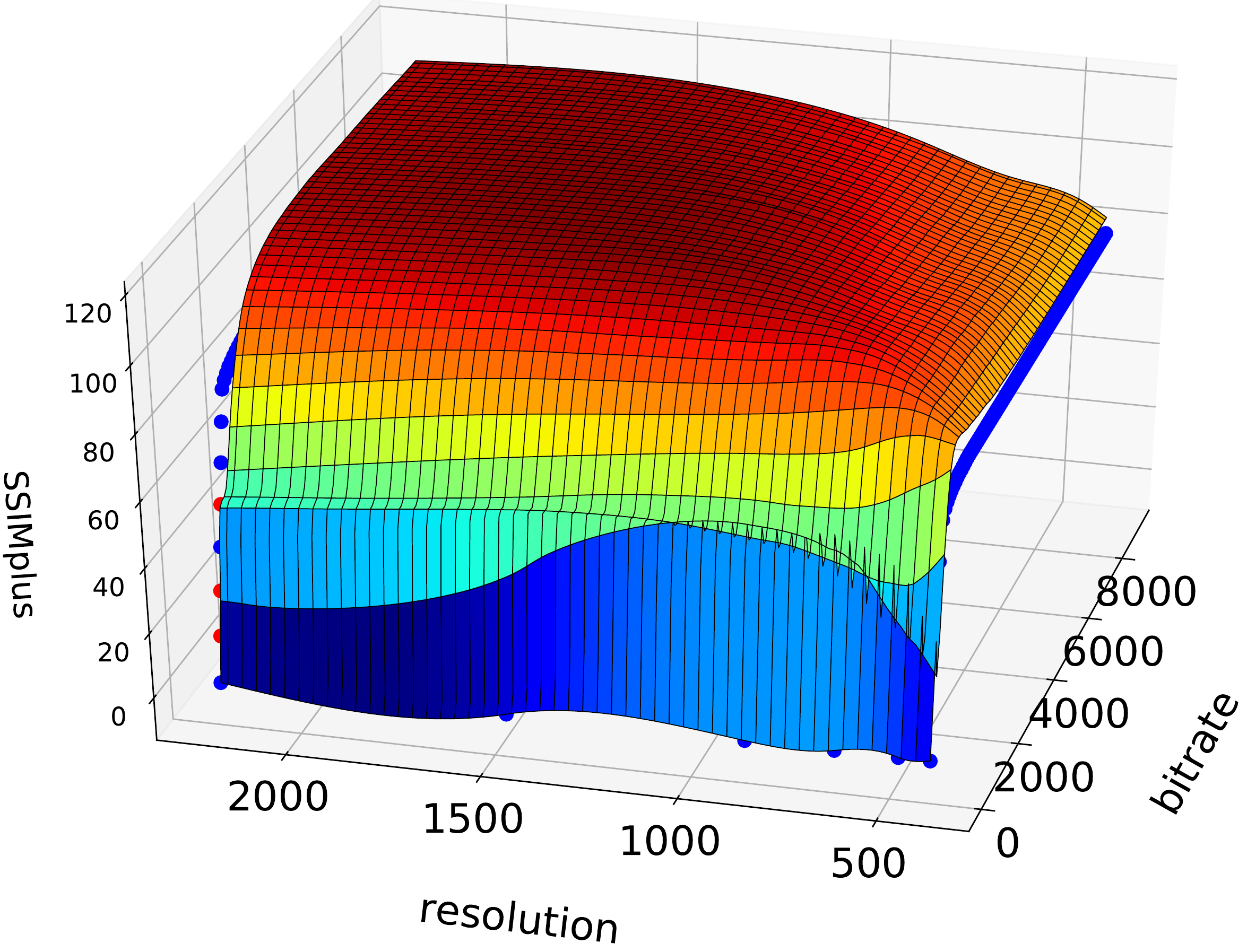}\label{subfig:no_mono}}
	\subfloat[]{\includegraphics[width=0.24\textwidth]{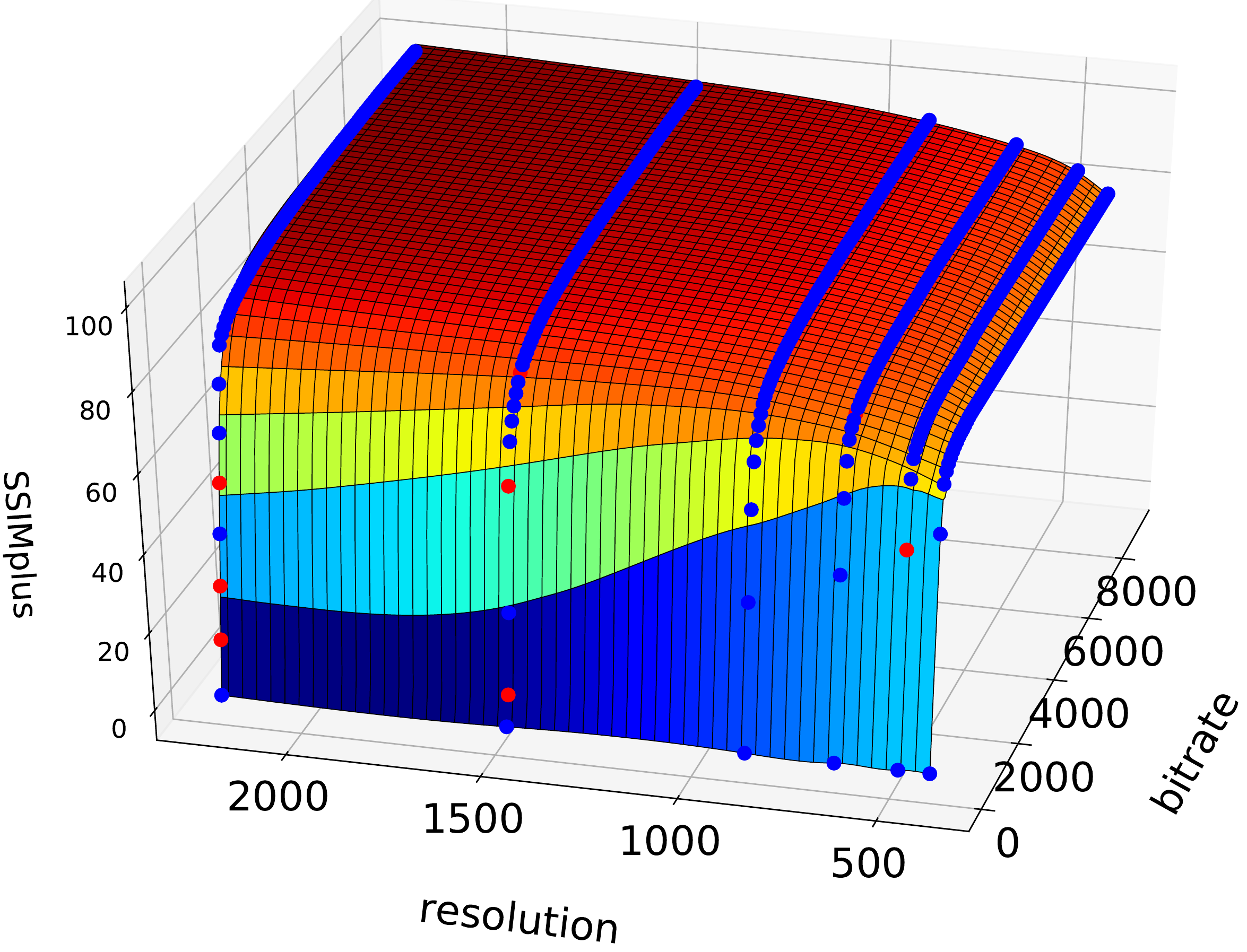}\label{subfig:with_mono}}
	\caption{The eGRD reconstructed surfaces (a) without and (b) with the monotonicity constraints. The points represent the discrete ground-truth GRD function. The red points indicate the training samples.}
	\label{fig:egrd_no_mono}
\end{figure}

To demonstrate the importance of the monotonicity assumption, we lift the constraints in Problem~\eqref{eq:proj}, and solve the system of linear equations with the least squares method.
The results are listed in Table~\ref{tab:egrd_no_mono}, from which we can see that  the robustness of eGRD deteriorates in general.
Particularly, the simplifed eGRD model easily overfits with more samples, which is also illustrated in Fig.~\ref{fig:egrd_no_mono}.
The reconstructed surface in Fig.~\ref{fig:egrd_no_mono} (a) severely violates the axial monotonicity of the GRD function. In comparison, the proposed eGRD method exploits the monotonicity constraints to regularize the shape of the reconstructed surface, leading to an accurate, smooth, and valid GRD prediction as shown in Fig.~\ref{fig:egrd_no_mono} (b).

\subsection{eGRD with Different Encoders}\label{subsec:exp_encoder}
\begin{table}[t!]
	\centering
	\caption{Performance of eGRD  training with H.264~\cite{ffmpegH264} and testing on VP9~\cite{ffmpegVP9}}
	\label{tab:egrd_h264_vp9}
	\begin{tabular}{c|cc|cc}
		\toprule
		\multirow{2}{*}{$N$/$S$} & \multicolumn{2}{c|}{RMSE} & \multicolumn{2}{c}{$l^\infty$ error}\\ \cline{2-5}
		  & Mean & Worst & Mean  & Worst \\
		\hline
		1 & 2.52 & 13.66 & 25.99 & 60.52 \\
		3 & 1.43 & 6.25 & 13.42 & 37.21 \\
		5 & 0.98 & 6.87 & 8.64 & 33.87 \\
		8 & 0.86 & 3.66 & 4.16 & 21.69 \\
		30 & 0.50 & 7.28 & 2.66 & 20.49 \\
		50 & 0.23 & 1.15 & 0.90 & 8.91 \\
		\bottomrule
	\end{tabular}
\end{table}
\begin{table}[t!]
	\centering
	\caption{Performance of eGRD   training with  VP9~\cite{ffmpegVP9} and testing on H.264~\cite{ffmpegH264}}
	\label{tab:egrd_vp9_h264}
	\begin{tabular}{c|cc|cc}
		\toprule
		\multirow{2}{*}{$N$/$S$} & \multicolumn{2}{c|}{RMSE} & \multicolumn{2}{c}{$l^\infty$ error}\\ \cline{2-5}
		 & Mean & Worst & Mean  & Worst \\
		\hline
		1 & 2.21 & 13.90 & 23.64 & 62.21 \\
		3 & 2.37 & 8.32 & 10.99 & 42.58 \\
		5 & 1.53 & 6.13 & 12.67 & 57.66 \\
		8 & 1.86 & 7.62 & 6.46 & 28.33 \\
		30 & 0.87 & 6.49 & 3.51 & 17.42 \\
		50 & 0.32 & 2.37 & 1.58 & 6.93 \\
		\bottomrule
	\end{tabular}
\end{table}

\begin{table}[t]
	\centering
	\caption{Mean and worst performance of eGRD on GRD functions measured by VMAF}
	\label{tab:egrd_vmaf}
	\begin{tabular}{c|cc|cc}
		\toprule
	\multirow{2}{*}{$N$/$S$}& \multicolumn{2}{c|}{RMSE} & \multicolumn{2}{c}{$l^\infty$ error}\\ \cline{2-5}
		 & Mean & Worst & Mean  & Worst \\
		\midrule
		1 & 6.67 & 20.45 & 25.33 & 69.12 \\
		3 & 2.58 & 13.51 & 12.12 & 48.73 \\
		5 & 1.28 & 7.24 & 10.90 & 55.47 \\
		8 & 0.86 & 6.32 & 6.10 & 29.96 \\
		30 & 0.34 & 1.75 & 3.38 & 17.06 \\
		50 & 0.24 & 1.12 & 3.53 & 11.93 \\
		\bottomrule
	\end{tabular}
\end{table}

The underlying principle behind eGRD applies to all rate-allocation strategies and encoding profiles as long as the four basic assumptions of GRD functions are satisfied. 
Here we investigate the generality of eGRD from one encoder to other encoders.
We split the Waterloo GRD database into two subsets, one containing $1,000$ GRD functions of the H.264 encoder~\cite{ffmpegH264}, and the other containing those of VP9~\cite{ffmpegVP9}.
We train eGRD on one subset, and test it on the other.
Tables~\ref{tab:egrd_h264_vp9} and \ref{tab:egrd_vp9_h264} encapsulate the cross-encoder performance, where we find that the reconstruction accuracy slightly declines when eGRD is trained and tested under different encoders.
This reflects the fact that each encoder makes respective assumptions on the distribution of video signal, and thus systematically deforms the GRD surface towards a specific direction.
Moreover, we are still able to achieve good reconstruction performance with sufficient samples, implying that the empirical GRD functions arising from the two encoders form almost the same space, though the learned principal components may differ.
This observation further enhances the practicality of the proposed eGRD method.
We believe the generalization capability arises from the similarity between the encoding processes of x264 and VP9. 
Nevertheless, for a substantially different encoder, it is safer to retrain eGRD on the new encoder before GRD reconstruction.
\begin{table*}[!t]
    \centering
    \caption{Average runtime in seconds of GRD models over $10,000$ runs. The time for video encoding and quality assessment is not included}
    \begin{tabular}{c|c c c c c c c c}
        \toprule
        Model & BD & Logistic & Logarithmic & Reciprocal & PCHIP & pGRD & tGRD & eGRD \\\midrule
        Training (s) & -- & -- & -- & -- & -- & -- & -- & 0.157 \\
        Testing (s) & 0.022 & 0.056 & 0.003 & 0.089 & 0.001 & 0.022 & 0.021 & 0.006 \\
        \bottomrule
    \end{tabular}
    \label{tab:runtime}
\end{table*}

\subsection{eGRD with Different VQA Models}\label{subsec:exp_vmaf}
The proposed eGRD algorithm does not attach itself to any specific VQA method.
To show this, we combine eGRD with another full-reference VQA model, VMAF~\cite{li2016VMAF}.
Again, we leverage the $1,080,000$ encoded video representations in the Waterloo GRD database, and evaluate perceptual quality using VMAF. From Table~\ref{tab:egrd_vmaf}, we see that the estimation accuracy and rate of convergence on the VMAF-based GRD functions are comparable to  SSIMplus-based ones, validating the generalizability of eGRD.
This can be partly ascribed to the fact that the mathematical assumptions made in Section~\ref{sec:theoretic} generally hold no matter which VQA model is employed.

\subsection{Model Complexity}
eGRD consists of a training phase and a testing phase.
During training, PCA is performed to calculate the eigenvectors, with a complexity of $\mathcal{O}(\min(M^3, K^3))$ for a data matrix $\mathbf{X} \in \mathbb{R}^{M \times K}$~\cite{johnstone2004sparse}, where $M$ and $K$ denote the number of observations and the dimension of each observation, respectively. 
On a computer with  3.60 GHz CPU and 16 GB RAM, it takes less than $1$ second to run PCA. 
The training of eGRD can be performed offline once, and will not affect the testing phase. 
Note that no training is needed for models with fixed basis.

In the testing phase, eGRD solves a quadratic programming problem, with a computational complexity of $\mathcal{O}(N^2)$ to 
$\mathcal{O}(N^3)$~\cite{stellato2017osqp,kouzoupis2018recent}, where $N$ denotes the number of parameters.
This complexity is similar to that of the reciprocal~\cite{toni2015optimal}, logarithmic~\cite{chen2016subjective}, and logistic~\cite{hanhart2014calculation} methods, which employ interior-point algorithms with a polynomial complexity~\cite{potra2000interior}.
In addition, both the BD~\cite{bjontegaard2001bda,bjontegaard2008bdb} and PCHIP~\cite{zern2010webm} models need to solve linear equation systems with a complexity of at most $O(MN^2 + N^3)$~\cite{thekerneltrip2018computational}.

We have also tested the runtime of all these methods, as reported in Table~\ref{tab:runtime}, from which we can observe that the proposed eGRD exhibits minimal overhead.

\subsection{Implications}
The theoretical foundation of eGRD has several important implications in real-world applications.
First, our mathematical analysis shows that all GRD functions live in a convex subset of an affine subspace, which assures a unique and error-bounded solution.
Second, the intrinsic low dimensionality of GRD functions may facilitate the use of eGRD in reduced-reference VQA.
Specifically, the coefficients of a few eigenvectors can be transmitted from the server to the client for  perceptual quality prediction.
The transmission of the coefficients may be economically favorable to raw quality scores when there are several encoding representations.
Third, the monotonicity properties of GRD functions reflect the trade-off between bitrate and perceptual quality, and thus are essential for subsequent video applications.

\section{Application of eGRD: Video Codec Comparison}
Video coding is the core technology in many modern video services.
In the past decades, new video compression algorithms keep springing up, claiming significant performance improvements over previous codecs. 
Algorithm~\ref{alg:codec_comparison} gives the general framework for video codec comparison.
Given a pair of codecs, Algorithm~\ref{alg:codec_comparison} first estimates the RD and distortion-rate (DR) curves of the two codecs from $S$ samples (typically $S = 4$), and then calculates the relative quality gain and bitrate saving between the two curves~\cite{bjontegaard2001bda,bjontegaard2008bdb,zern2010webm,hanhart2014calculation}.
The reliability of codec comparison depends heavily on the RD/DR function estimation method.

\begin{figure}[!t]
	\removelatexerror
	\begin{algorithm}[H]
		\caption{General Framework for Video Codec Comparison }
		\label{alg:codec_comparison}
		\SetKwFunction{VQA}{VQA}\SetKwFunction{Mean}{Mean}
		\SetKwInOut{Input}{input}\SetKwInOut{Output}{output}
		
		\Input{Two codecs $A$ and $B$; A set of source videos $\mathcal{V} = \{v_i\}_{i=1}^{M}$; A set of target encoding bitrates $\{x_k\}_{k=1}^S$}
		\Output{Average quality gain $\Delta Q$; Average bitrate saving $\Delta R$}
		\BlankLine
		
		\For{$i \leftarrow 1$ \KwTo $M$}
		{
			\For{$k \leftarrow 1$ \KwTo $S$}{
				$v_{i, k}^A \leftarrow$ Encode $v_i$ with $A$ at $x_k$\;
				$z_{i, k}^A \leftarrow$ \VQA{$v_{i, k}^A$}\;
				$\hat{x}_{i, k}^A \leftarrow$ Log of actual bitrate of $v_{i, k}^A$\;
				$v_{i, k}^B \leftarrow$ Encode $v_i$ with $B$ at $x_k$\;
				$z_{i, k}^B \leftarrow$ \VQA{$v_{i, k}^B$}\;
				$\hat{x}_{i, k}^B \leftarrow$ Log of actual bitrate of $v_{i, k}^B$\;
			}
			
			Fit the rate-distortion (RD) function $f_i^A$ of Codec $A$ from $\{(\hat{x}_{i, k}^A, z_{i, k}^A)\}_{k=1}^S$\;
			Fit the distortion-rate (DR) function $g_i^A$ of Codec $A$ from $\{(z_{i, k}^A, \hat{x}_{i, k}^A)\}_{k=1}^S$\;
			Fit the RD function $f_i^B$ of Codec $B$ from $\{(\hat{x}_{i, k}^B, z_{i, k}^B)\}_{k=1}^S$\;
			Fit the DR function $g_i^B$ of Codec $B$ from $\{(z_{i, k}^B, \hat{x}_{i, k}^B)\}_{k=1}^S$\;
			\BlankLine
			$\hat{x}_{i, L} \leftarrow \max[\min(\hat{x}_{i, 1}^A, \ldots, \hat{x}_{i, S}^A), \min(\hat{x}_{i, 1}^B, \ldots, \hat{x}_{i, S}^B)]$\;
			$\hat{x}_{i, H} \leftarrow \min[\max(\hat{x}_{i, 1}^A, \ldots, \hat{x}_{i, S}^A), \max(\hat{x}_{i, 1}^B, \ldots, \hat{x}_{i, S}^B)]$\;
			$\Delta Q_i \leftarrow \frac{1}{\hat{x}_{i, H} - \hat{x}_{i, L}}\int_{\hat{x}_{i, L}}^{\hat{x}_{i, H}} \left[f_i^B(\hat{x})-f_i^A(\hat{x})\right]d\hat{x}$\;
			\BlankLine
			$z_{i, L} \leftarrow \max[\min(z_{i, 1}^A, \ldots, z_{i, S}^A), \min(z_{i, 1}^B, \ldots, z_{i, S}^B)]$\;
			$z_{i, H} \leftarrow \min[\max(z_{i, 1}^A, \ldots, z_{i, S}^A), \max(z_{i, 1}^B, \ldots, z_{i, S}^B)]$\;
			\nl$\Delta R_i \leftarrow 10^{\frac{1}{z_{i, H} - z_{i, L}}\int_{z_{i, L}}^{z_{i, H}} \left[g_i^B(z)-g_i^A(z)\right]dz} - 1$\label{eq:log_rsaving}\;
		}
		$\Delta Q \leftarrow$ \Mean{$\Delta Q_1, \ldots, \Delta Q_M$}\;
		$\Delta R \leftarrow$ \Mean{$\Delta R_1, \ldots, \Delta R_M$}.
	\end{algorithm}
\end{figure}

Since eGRD can accurately estimate a 2D GRD function from very few samples, we adopt it for 1D RD function estimation, and introduce an eGRD-based video codec comparison method. We generally follow Algorithm~\ref{alg:codec_comparison} with several modifications.
First, the original eGRD method only gives a discrete RD function.
We estimate a continuous function $f$ by linear interpolation.
Second, we obtain the DR function $g$ by taking the inverse of $f$.
Third, we improve the calculation of $\Delta R_i$ by strictly following the definition instead of using the inaccurate approximation at Line~\ref{eq:log_rsaving} in Algorithm~\ref{alg:codec_comparison}:
\begin{equation}
\label{eq:r_saving_egrd} \Delta R_i \leftarrow \frac{1}{z_{i, H} - z_{i, L}}\int_{z_{i, L}}^{z_{i, H}} \left[\frac{g_i^B(z)-g_i^A(z)}{g_i^A(z)}\right]dz.
\end{equation}
This is made possible because  eGRD directly estimates the bitrate $x$ rather than its logarithm $\hat{x}$.

\subsection{Local and Global Codec Comparison}

We compare the proposed video codec comparison method with the logistic model~\cite{hanhart2014calculation}, BD~\cite{bjontegaard2001bda,bjontegaard2008bdb}, and  PCHIP~\cite{zern2010webm}.
Unlike our model, all competing methods work with log bitrate as indicated in Algorithm \ref{alg:codec_comparison}.
The widely-used BD model adopts cubic polynomials, while PCHIP employs Hermite interpolating polynomials to fit RD/DR functions.
More recently, a logistic model~\cite{hanhart2014calculation} is proposed to fit the RD function, with an analytical inverse as the corresponding DR function. In this experiment, we consider two practical video codecs, x264~\cite{ffmpegH264} and VP9~\cite{ffmpegVP9} at the resolution of $1,920 \times 1,080$, based on the videos from the Waterloo GRD database.
To quantify the performance, we first calculate the ground-truth quality gain and bitrate saving on every test video using the densely-sampled RD/DR functions provided in the Waterloo GRD database.
Then we estimate quality gains and bitrate savings using these codec comparison models.
The average errors over a set of test videos are reported.


\begin{table*}[!t]
    \centering
    \caption{Average absolute estimation error of $\Delta Q$ and $\Delta R$ in  local codec comparison}
    \label{tab:codec_comp_local}
    \begin{tabular}{c|c c c c|c c c c}
        \toprule
        & \multicolumn{4}{c|}{Error of $\Delta Q$} & \multicolumn{4}{c}{Error of $\Delta R$} \\\cline{2-9}
        & BD & Logistic & PCHIP & eGRD & BD & Logistic & PCHIP & eGRD \\\midrule
        Uniform Sampling& 0.772 & 0.648 & 0.763 & \textbf{0.512} & $3.616 \times 10^{32}$ & 3.267 & 4.086 & \textbf{2.155} \\
        Uncertainty Sampling& 0.958 & 0.824 & 0.835 & \textbf{0.529} & 4.255 & 3.812 & 4.211 & \textbf{2.388} \\
        \bottomrule
    \end{tabular}
\end{table*}

\begin{table*}[!t]
    \centering
    \caption{Average absolute estimation error of $\Delta Q$ and $\Delta R$ in global codec comparison}
    \label{tab:codec_comp_global}
    \begin{tabular}{c|c c c c|c c c c}
        \toprule
        & \multicolumn{4}{c|}{Error of $\Delta Q$} & \multicolumn{4}{c}{Error of $\Delta R$} \\\cline{2-9}
        & BD & Logistic & PCHIP & eGRD & BD & Logistic & PCHIP & eGRD \\\midrule
        Uniform Sampling& 4.101 & 0.408 & 1.100 & \textbf{0.390} & $3.723 \times 10^{32}$ & 3.587 & 8.629 & \textbf{2.019} \\
        Uncertainty Sampling& 3.656 & 0.514 & 1.239 & \textbf{0.403} & $4.374 \times 10^{76}$ & 4.079 & 9.162 & \textbf{2.193} \\
        \bottomrule
    \end{tabular}
\end{table*}

The RD sample set $\{x_k\}_{k=1}^S$ is a critical aspect of reliable codec comparison.
Unfortunately, to the best of our knowledge, there are no widely accepted querying algorithms in codec comparison.
Therefore, we investigate two sampling strategies - uniform sampling in the log bitrate scale and uncertainty sampling~\cite{duanmu2019modeling}.
Uniform and uncertainty sampling strategies produce the querying bitrate sets $\{100, 300, 900, 2800\}$ and $\{100, 300, 1100, 2800\}$, respectively, which closely resemble bitrate selection in many video codec comparison studies~\cite{ohm2012hevc,li2019avc,cock2016comparison}. We then compute quality gains and bitrate savings on the bitrate interval $[100, 2800]$ kbps without examining the extrapolation capability of competition models.

For our eGRD-based codec comparison model, we randomly select $1,600$ RD functions from $800$ source videos in the Waterloo GRD database for training, and leave the rest $200$ videos for testing.
We repeat the random splitting $50$ times, and report the median results in Table~\ref{tab:codec_comp_local}. We find  that the proposed eGRD-based method achieves the lowest estimation errors in both quality gains and bitrate savings.
Note that all competing models generally follow the same framework in Algorithm~\ref{alg:codec_comparison}, implying that the use of eGRD is the main reason for the performance improvement.

In practice, video engineers are not only interested in the codec comparison at a particular interval, but also across all bitrates.
Here we examine the performance of codec comparison models in the full bitrate range (\textit{i.e.}, $[0, 9000]$ kbps)  given only RD samples on the local bitrate interval $[x_{i, L}, x_{i, H}]$.
Table~\ref{tab:codec_comp_global} shows the results, where we observe that the eGRD-based codec comparison model achieves a more significant improvement, thanks to the wide operating range of eGRD.

\subsection{Discussion}
\begin{figure}[p!t]
	\centering
	\subfloat[]{\includegraphics[width=0.24\textwidth]{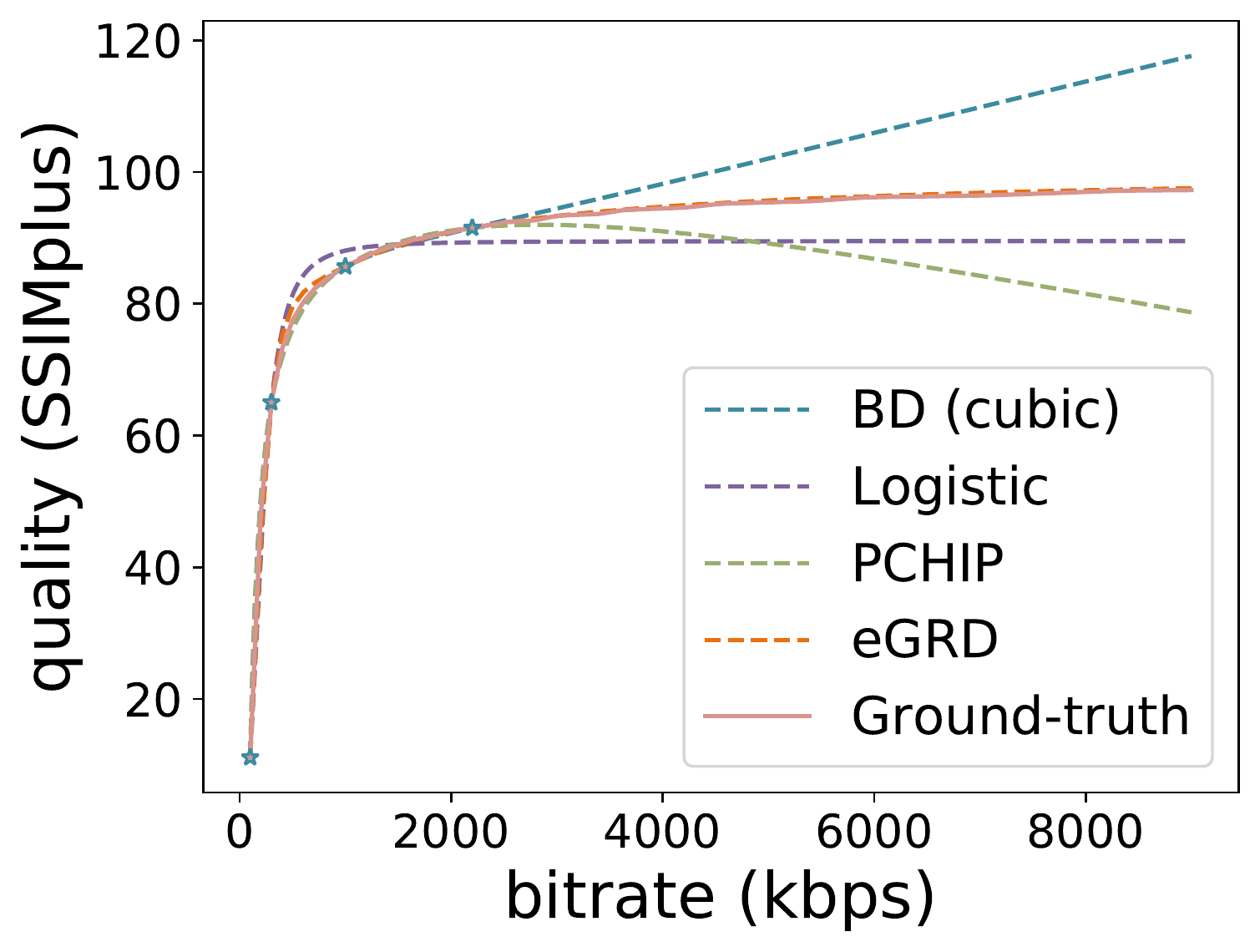}\label{subfig:video94}}
	\subfloat[]{\includegraphics[width=0.24\textwidth]{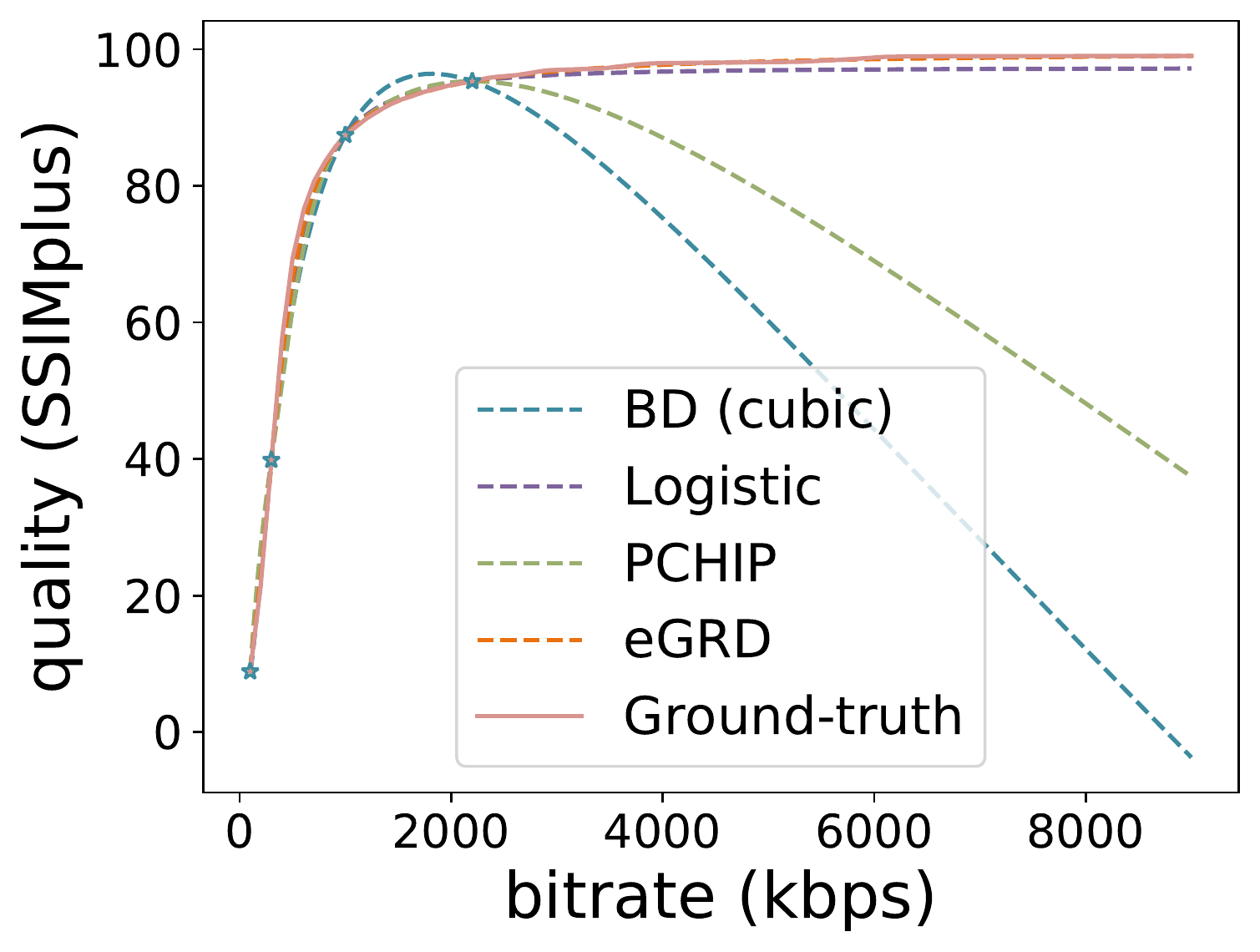}\label{subfig:video9}}
	\caption{Comparison of RD curve estimations. In both figures, the eGRD model gives the best approximations, while the other three models can significantly diverge from the ground-truth.}
	\label{fig:estimation_acc}
\end{figure}

\begin{figure}[p!t]
	\centering
	\subfloat[]{\includegraphics[width=0.24\textwidth]{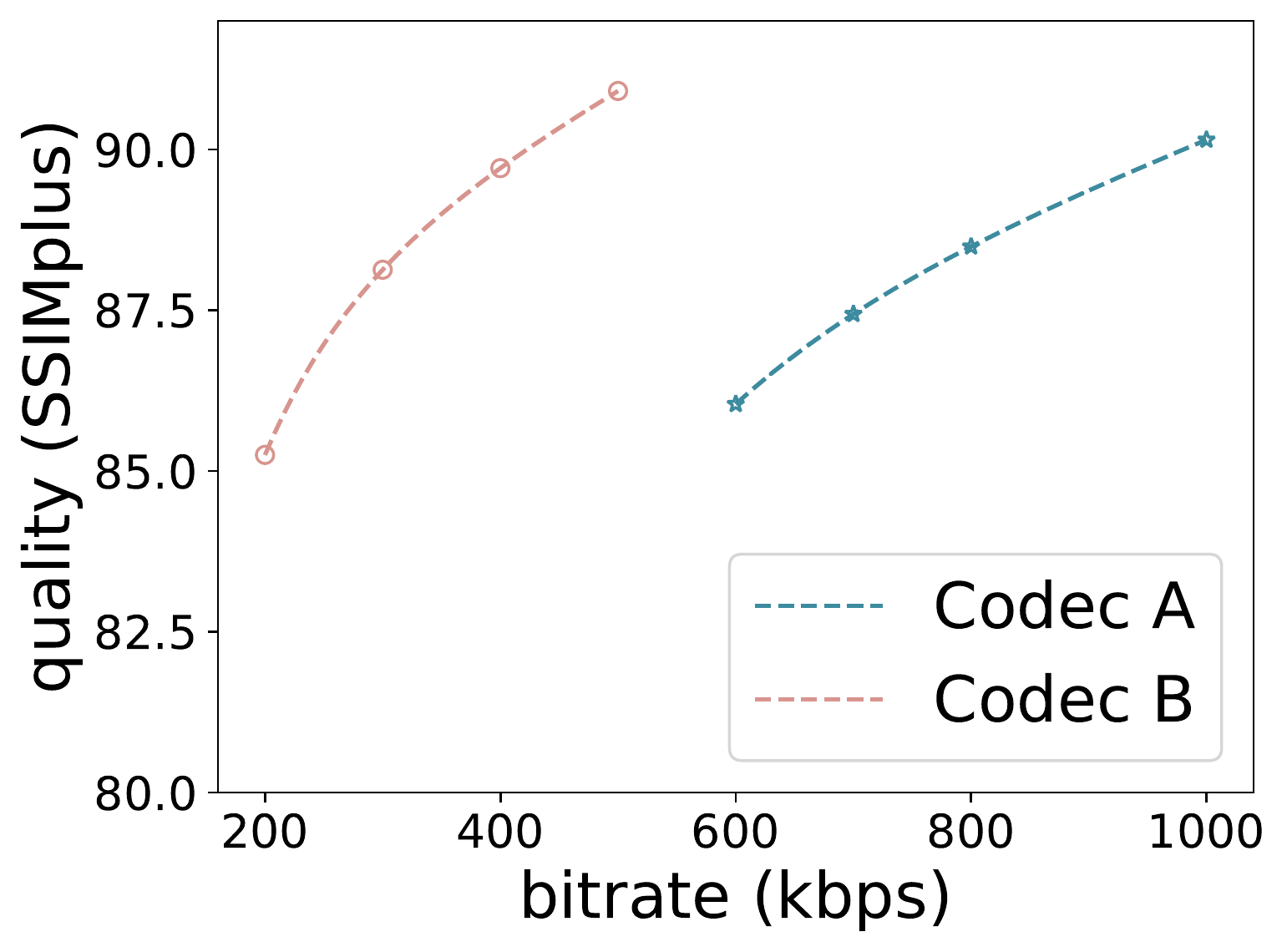}\label{subfig:no_quality_gain}}
	\subfloat[]{\includegraphics[width=0.24\textwidth]{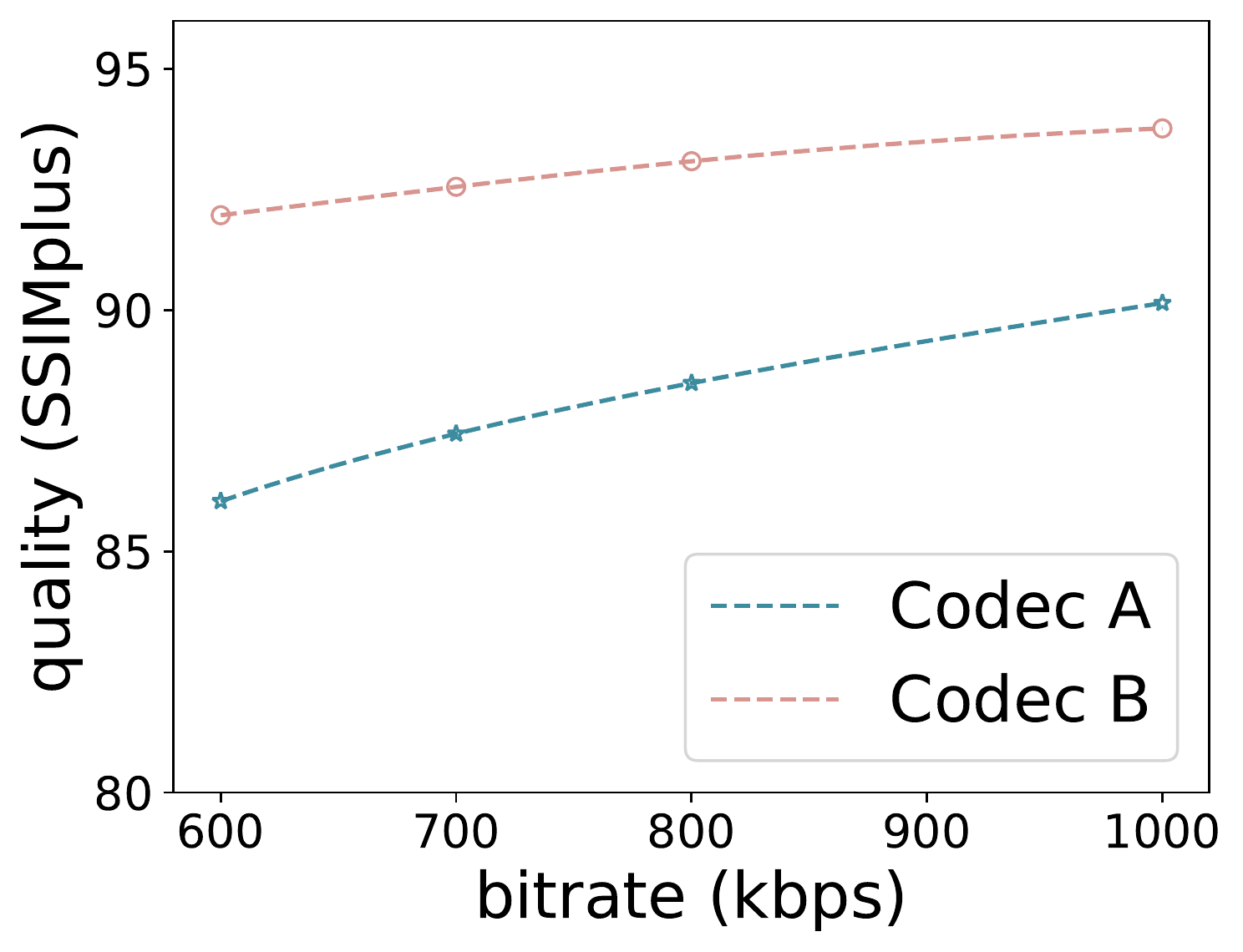}\label{subfig:no_bitrate_saving}}
	\caption{Existing video codec comparison models often fail when either (a) the bitrate range or (b) the quality range of the sample representations of one encoder does not overlap with that of the other.}
	\label{fig:codec_comparison_failure}
\end{figure}

To gain an intuition on how the eGRD-based model outperforms the other three, we select two real-world examples from the Waterloo GRD database, and draw their respective estimated RD curves in Fig.~\ref{fig:estimation_acc}, where we find that eGRD and PCHIP perform well in the bitrate range between the two furthest sample points.
By contrast, the logistic model gives an inaccurate estimate in Fig.~\ref{fig:estimation_acc} (a), and  BD even produces a non-monotonic RD curve in Fig.~\ref{fig:estimation_acc} (b).
When it comes to the bitrate range that requires extrapolation, we find that only eGRD is able to accurately predict the ground-truth RD curves.
Due to the lack of regularizers,  BD and PCHIP may not be able to reconstruct valid RD curves with  limited samples. Although the logistic model can produce a valid RD curve, it saturates too early, failing to reflect the quality gains at high bitrates.
The inaccurate extrapolation explains why the three existing models~\cite{bjontegaard2001bda,bjontegaard2008bdb,zern2010webm,hanhart2014calculation} coincidentally restrict their quality gains and bitrate savings in the domain covered by the samples.
However, such restrictions may cause severe problems in practice.
Fig.~\ref{fig:codec_comparison_failure} illustrates two real-world examples where estimation of either  quality gains or bitrate savings fails.

\begin{figure}[p!t]
	\centering
	\subfloat[]{\includegraphics[width=0.24\textwidth]{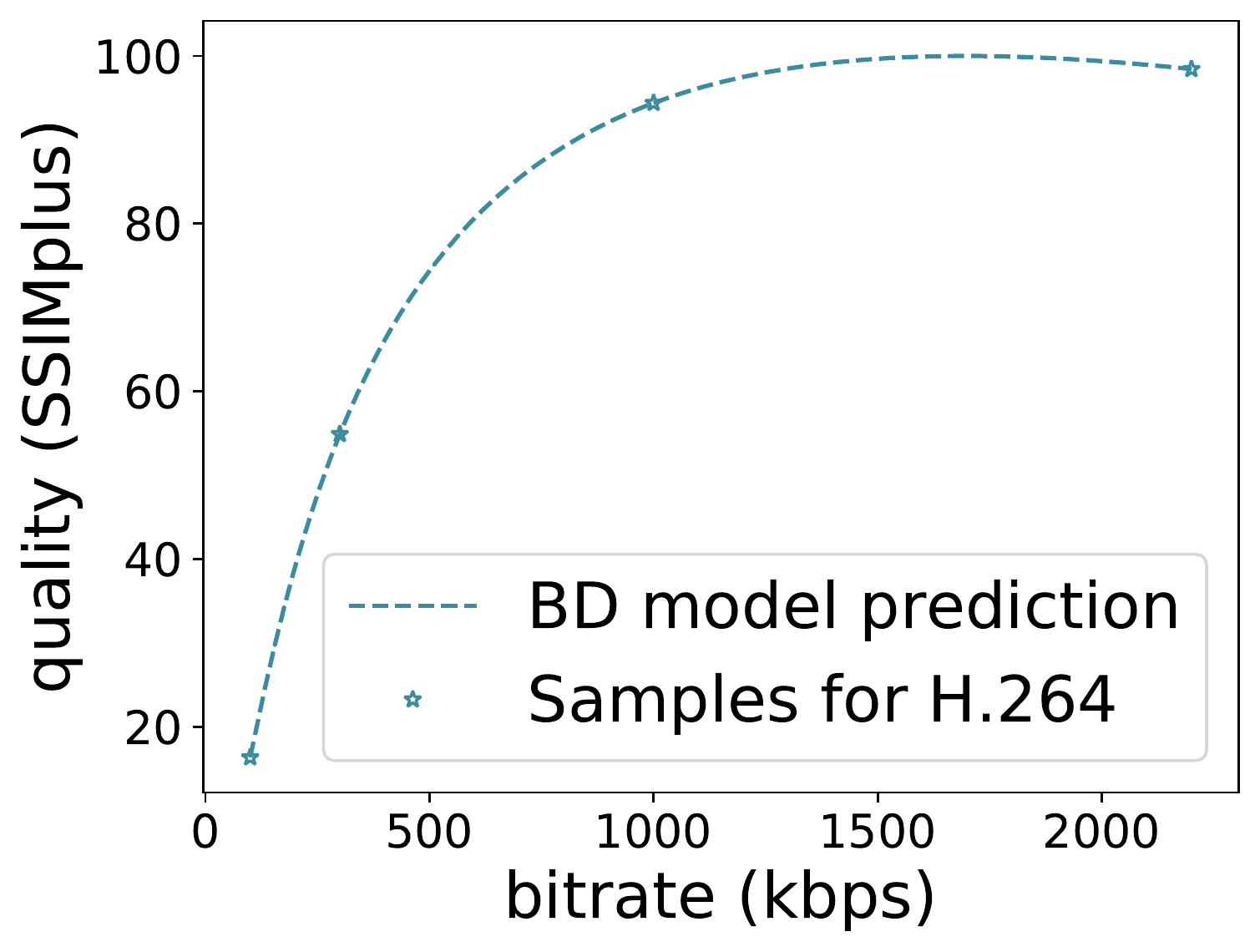}\label{subfig:non_mono_rd}}
	\subfloat[]{\includegraphics[width=0.24\textwidth]{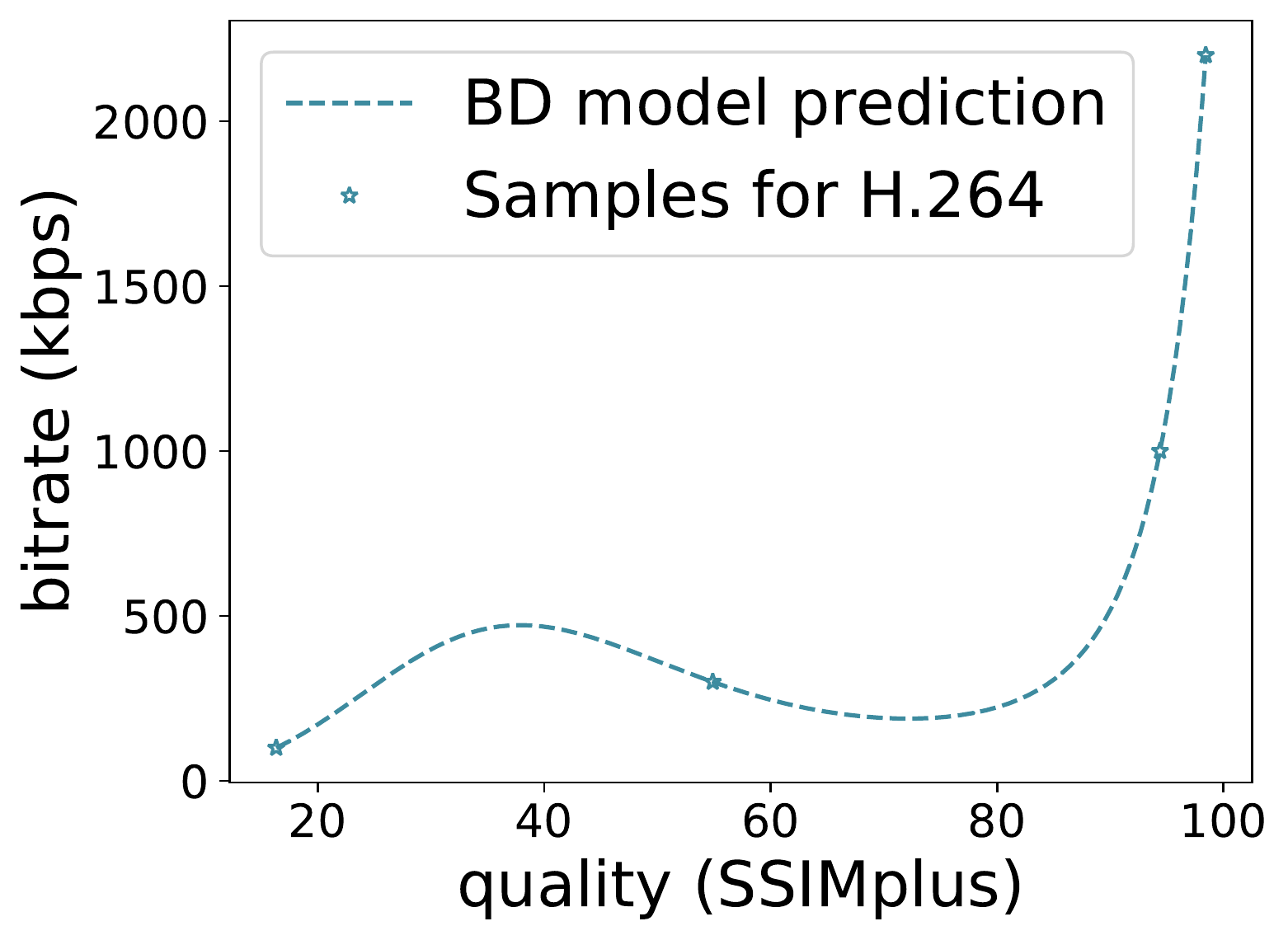}\label{subfig:non_mono_dr}}
	\caption{Non-monotonic (a) RD and (b) DR functions of the same video fitted by the BD model.}
	\label{fig:non_mono}
\end{figure}

\begin{figure}[p!t]
	\centering
	\subfloat[]{\includegraphics[width=0.24\textwidth]{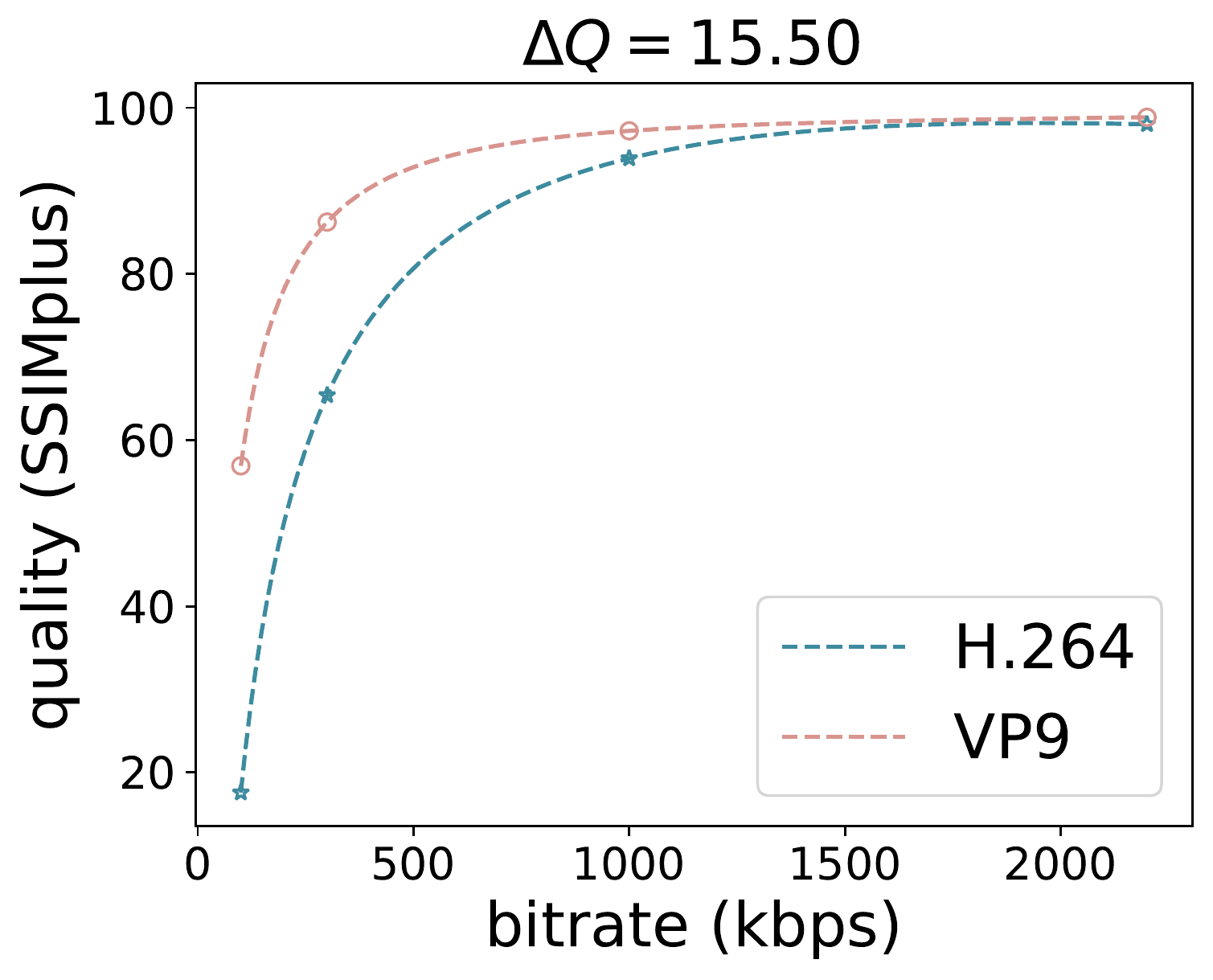}\label{subfig:q_gain}}
	\subfloat[]{\includegraphics[width=0.24\textwidth]{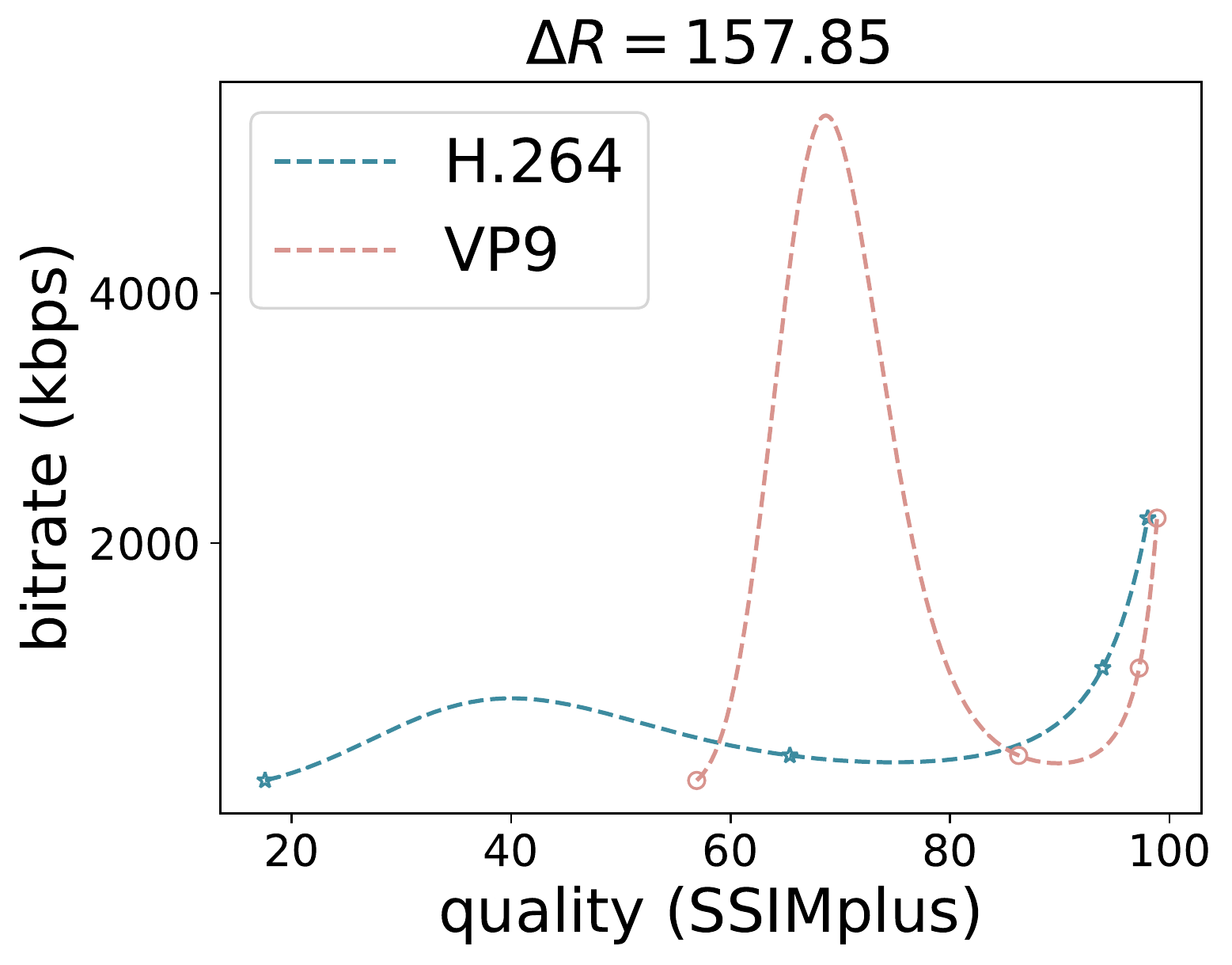}\label{subfig:r_saving}}
	\caption{A real-world counterexample of BD. VP9~\cite{ffmpegVP9} is compared to H.264~\cite{ffmpegH264} on the same video content. The positive $\Delta Q$ in (a) indicates that VP9 outperforms H.264, while the positive $\Delta R$ in (b) indicates the opposite.}
	\label{fig:q_r_conflict}
\end{figure}

Until now, the BD model is still the most prevalent tool to compare the performance of two video codecs~\cite{tan2016video,grois2017performance,akyazi2018comparison,li2019avc}.
However, in the above experiments,  BD  performs the worst.
By scrutinizing the experimental results more carefully, we find two more serious problems of BD.
First, it frequently produces non-monotonic RD/DR curves, even though the given samples are monotonic (see Fig.~\ref{fig:non_mono}).
Second, BD fits RD and DR functions independently, so the two functions may not be the inverse of each other.
As a result, the quality gain $\Delta Q$ and the bitrate saving $\Delta R$ resulting from BD may sometimes contradict each other (see
Fig.~\ref{fig:q_r_conflict}).

\section{Conclusion}\label{sec:conclusion}
GRD functions provide a comprehensive description of the relationship between the encoding profile and perceptual quality, based on which many video-related applications are made possible.
In this work, we propose a general GRD model for accurate GRD function reconstruction from sparse samples.
The performance improvements of our model may arise from the data-driven eigen basis for representing  real-world GRD functions and  the axial monotonicity constraints for preventing the model from overfitting. 
Extensive experiments on the Waterloo GRD database have shown that the proposed eGRD algorithm is able to accurately reconstruct GRD functions with a very small number of samples, and is robust in various practical scenarios.

%

\appendices
\section{Encoding Configurations}\label{sec:encoding_cmd}
Considering video-on-demand adaptive streaming as one of the major application scenarios, we generally follow the encoding strategies as suggested in~\cite{stream2015dash}.
We employ the open-source FFmpeg software to encode the videos. 
The $\text{x264}$~\cite{ffmpegH264} and $\text{vpx-vp9}$~\cite{ffmpegVP9} libraries are used for H.264 and VP9 encoding, respectively.
The detailed specifications are shown in Table~\ref{tab:encoding_spec}.

\begin{table*}[!t]
    \centering
    \caption{Encoding configurations}
    \label{tab:encoding_spec}
    \begin{tabular}{c|c}
        \toprule
        Codec & Command \\
        \midrule
        \multirow{4}{*}{H.264} & ffmpeg -y -r framerate -f rawvideo -pix\_fmt yuv420p -s 1920x1080 -i ref.yuv -c:v libx264 -preset medium\\
        & -s target\_resolution -r framerate -b:v target\_bitrate -pass 1 -f mp4 /dev/null;\\
        & ffmpeg -r framerate -f rawvideo -pix\_fmt yuv420p -s 1920x1080 -i ref.yuv -c:v libx264 -preset medium \\
        & -s target\_resolution -r framerate -b:v target\_bitrate -pass 2 out.mp4 \\
        \hline
        \multirow{4}{*}{VP9} & ffmpeg -y -r framerate -f rawvideo -pix\_fmt yuv420p -s 1920x1080 -i ref.yuv -c:v libvpx-vp9 -speed 1\\
        & -s target\_resolution -r framerate -b:v target\_bitrate -pass 1 -f webm /dev/null;\\
        & ffmpeg -r framerate -f rawvideo -pix\_fmt yuv420p -s 1920x1080 -i ref.yuv -c:v libvpx-vp9 -speed 1 \\
        & -s target\_resolution -r framerate -b:v target\_bitrate -pass 2 out.webm \\
         \bottomrule
    \end{tabular}
\end{table*}

\section{Uncertainty Sampling}\label{sec:uncertainty}
All GRD models rely on a set of attribute-quality pairs.
The ``importance'' of such pairs in  GRD function reconstruction could be drastically different.
The uncertainty sampling proposed in~\cite{duanmu2019modeling} formalizes this intuition based on  information theory.
Here we provide a brief description of the method.
Assuming $\mathbf{f}$ follows a multivariate Gaussian distribution with a covariance matrix $\boldsymbol{\Sigma}$, the total uncertainty of $\mathbf{f}$ is characterized by its joint entropy.
If $\mathbf{f}$ is divided into two parts such that ${\bf f} = \left[ {\begin{array}{*{20}{c}}
{\bf f}_1\\
{\bf f}_2
\end{array}} \right]$ and $\boldsymbol{\Sigma} = \begin{bmatrix}\boldsymbol{\Sigma}_{11} & \boldsymbol{\Sigma}_{12} \\\boldsymbol{\Sigma}_{21} & \boldsymbol{\Sigma}_{22} \end{bmatrix}$, and ${\bf f}_2$ is observed (\textit{e.g.}, ${\bf f}_2={\bf a}$), then the remaining uncertainty is given by the conditional entropy
\begin{align*}
&H_{{\bf f}_1|{\bf f}_2}({\bf f}_1|{\bf f}_2) = \frac{1}{2}\log(|\boldsymbol{\bar{\Sigma}}|) + \mathrm{const},
\end{align*}
where
\begin{align*}
&\boldsymbol{\bar{\Sigma}} = \boldsymbol{\Sigma}_{11} - \boldsymbol{\Sigma}_{12}\boldsymbol{\Sigma}_{22}^{-1}\boldsymbol{\Sigma}_{21}.
\end{align*}
The most informative sample is found by minimizing the log determinant of the conditional covariance matrix~\cite{bishop2006pattern}
\begin{equation*}
\begin{aligned}
& \underset{i}{\text{min}}
& \log |\boldsymbol{\bar{\Sigma}}_i| = \underset{i}{\text{min}}
& & \log\left\vert\boldsymbol{{\Sigma}}_{\setminus ii} - \frac{\boldsymbol{{\sigma}}_{\setminus i}\boldsymbol{{\sigma}}_{ \setminus i}^T}{{\sigma}_{ii}}\right\vert,
\end{aligned}
\end{equation*}
where $\boldsymbol{{\Sigma}}_{\setminus ii}$ is a submatrix of  $\boldsymbol{{\Sigma}}$ excluding the $i$-th row and the $i$-th column, $\boldsymbol{{\sigma}}_{\setminus i}$ is the $i$-th column of $\boldsymbol{{\Sigma}}$ excluding the $i$-th entry, and $\sigma_{ii}$ is the $i$-th diagonal element of $\boldsymbol{{\Sigma}}$, respectively.
Minimizing the conditional entropy directly is computationally expensive, especially when the dimensionality is high. Alternatively, we minimize a modified upper bound of the conditional entropy:
\begin{equation*}
\begin{aligned}
& \underset{i}{\text{min}}
& & \mathrm{tr}\left(\boldsymbol{{\Sigma}}_{\setminus ii} - \frac{\boldsymbol{{\sigma}}_{\setminus i}\boldsymbol{{\sigma}}_{ \setminus i}^T}{{\sigma}_{ii}}\right),
\end{aligned}
\end{equation*}
where we make use of 
\begin{equation*}
\log\left\vert\boldsymbol{{\Sigma}}_{\setminus ii} - \frac{\boldsymbol{{\sigma}}_{\setminus i}\boldsymbol{{\sigma}}_{ \setminus i}^T}{{\sigma}_{ii}}\right\vert \leq \mathrm{tr}\left(\boldsymbol{{\Sigma}}_{\setminus ii} - \frac{\boldsymbol{{\sigma}}_{\setminus i}\boldsymbol{{\sigma}}_{ \setminus i}^T}{{\sigma}_{ii}}-{\bf I} \right)
\end{equation*}
and ${\bf I}$ is the identity matrix.
This process is applied iteratively, resulting in a sequence of optimal samples in terms of uncertainty reduction.
Note that the new covariance matrix $\boldsymbol{\bar{\Sigma}}$ does not depend on any specific observation of ${\bf f}_2$. As a result, the algorithm produces the same sampling sequence for all GRD functions.






%
\bibliographystyle{IEEEtran}
\bibliography{ref} 

\begin{thebibliography}{10}
\providecommand{\url}[1]{#1}
\csname url@samestyle\endcsname
\providecommand{\newblock}{\relax}
\providecommand{\bibinfo}[2]{#2}
\providecommand{\BIBentrySTDinterwordspacing}{\spaceskip=0pt\relax}
\providecommand{\BIBentryALTinterwordstretchfactor}{4}
\providecommand{\BIBentryALTinterwordspacing}{\spaceskip=\fontdimen2\font plus
\BIBentryALTinterwordstretchfactor\fontdimen3\font minus
  \fontdimen4\font\relax}
\providecommand{\BIBforeignlanguage}[2]{{%
\expandafter\ifx\csname l@#1\endcsname\relax
\typeout{** WARNING: IEEEtran.bst: No hyphenation pattern has been}%
\typeout{** loaded for the language `#1'. Using the pattern for}%
\typeout{** the default language instead.}%
\else
\language=\csname l@#1\endcsname
\fi
#2}}
\providecommand{\BIBdecl}{\relax}
\BIBdecl

\bibitem{berger1975rate}
T.~Berger, ``Rate distortion theory and data compression,'' in \emph{Advances
  in Source Coding}, 1975, pp. 1--39.

\bibitem{shannon1959coding}
C.~E. Shannon, ``Coding theorems for a discrete source with a fidelity
  criterion,'' \emph{Institute of Radio Engineers, International Convention
  Record}, vol.~4, no.~4, pp. 142--163, Mar. 1959.

\bibitem{grois2013performance}
D.~Grois, D.~Marpe, A.~Mulayoff, B.~Itzhaky, and O.~Hadar, ``Performance
  comparison of {H.265/MPEG-HEVC}, {VP9}, and {H.264/MPEG-AVC} encoders,'' in
  \emph{Picture Coding Symposium}, 2013, pp. 394--397.

\bibitem{bitmovin2018codec}
\BIBentryALTinterwordspacing
F.~Christian. (2018) Multi-codec {DASH} dataset: {An} evaluation of {AV1},
  {AVC}, {HEVC} and {VP9}. [Online]. Available:
  \url{https://bitmovin.com/av1-multi-codec-dash-dataset/}
\BIBentrySTDinterwordspacing

\bibitem{wang2012ssim}
S.~Wang, A.~Rehman, Z.~Wang, S.~Ma, and W.~Gao, ``{SSIM}-motivated
  rate-distortion optimization for video coding,'' \emph{IEEE Transactions on
  Circuits and Systems for Video Technology}, vol.~22, no.~4, pp. 516--529,
  Apr. 2012.

\bibitem{ou2014q}
Y.~Ou, Y.~Xue, and Y.~Wang, ``{Q-STAR}: {A} perceptual video quality model
  considering impact of spatial, temporal, and amplitude resolutions,''
  \emph{IEEE Transactions on Image Processing}, vol.~23, no.~6, pp. 2473--2486,
  Jun. 2014.

\bibitem{zhang2013qoe}
W.~Zhang, Y.~Wen, Z.~Chen, and A.~Khisti, ``{QoE}-driven cache management for
  {HTTP} adaptive bit rate streaming over wireless networks,'' \emph{IEEE
  Transactions on Multimedia}, vol.~15, no.~6, pp. 1431--1445, Oct. 2013.

\bibitem{toni2015optimal}
L.~Toni, R.~Aparicio-Pardo, K.~Pires, G.~Simon, A.~Blanc, and P.~Frossard,
  ``Optimal selection of adaptive streaming representations,'' \emph{ACM
  Transactions on Multimedia Computing, Communications, and Applications},
  vol.~11, no.~2, pp. 1--43, Feb. 2015.

\bibitem{de2016complexity}
J.~De~Cock, Z.~Li, M.~Manohara, and A.~Aaron, ``Complexity-based
  consistent-quality encoding in the cloud,'' in \emph{IEEE International
  Conference on Image Processing}, 2016, pp. 1484--1488.

\bibitem{chen2016subjective}
C.~Chen, S.~Inguva, A.~Rankin, and A.~Kokaram, ``A subjective study for the
  design of multi-resolution {ABR} video streams with the {VP9} codec,'' in
  \emph{Electronic Imaging}, 2016, pp. 1--5.

\bibitem{wang2015objective}
Z.~Wang, K.~Zeng, A.~Rehman, H.~Yeganeh, and S.~Wang, ``Objective video
  presentation {QoE} predictor for smart adaptive video streaming,'' in
  \emph{Proceedings of SPIE Optical Engineering+Applications}, 2015, pp.
  95\,990Y.1--95\,990Y.13.

\bibitem{chen2017encoding}
C.~Chen, Y.~Lin, A.~Kokaram, and S.~Benting, ``Encoding bitrate optimization
  using playback statistics for {HTTP}-based adaptive video streaming,''
  \emph{arXiv preprint arXiv:1709.08763}, Sep. 2017.

\bibitem{bjontegaard2001bda}
G.~Bj{\o}ntegaard, ``Calculation of average {PSNR} differences between
  rd-curves,'' Video Coding Experts Group (VCEG), Austin, TX, USA, Tech. Rep.
  VCEG-M33, ITU-T SG 16/Q6, 13th VCEG Meeting, Apr. 2001.

\bibitem{bjontegaard2008bdb}
------, ``Improvements of the {BD-PSNR} model,'' Video Coding Experts Group
  (VCEG), Berlin, Germany, Tech. Rep. VCEG-AI11, ITU-T SG 16/Q6, 35th VCEG
  Meeting, Jul. 2008.

\bibitem{duanmu2019modeling}
Z.~Duanmu, W.~Liu, and Z.~Wang, ``Modeling generalized rate-distortion
  functions,'' \emph{arXiv preprint arXiv:1906.05178}, Jun. 2019.

\bibitem{kreuzberger2016comparative}
C.~Kreuzberger, B.~Rainer, H.~Hellwagner, L.~Toni, and P.~Frossard, ``A
  comparative study of {DASH} representation sets using real user
  characteristics,'' in \emph{International Workshop on Network and OS Support
  for Digital Audio and Video}, 2016, pp. 1--4.

\bibitem{ISO2012Dash}
\BIBentryALTinterwordspacing
{DASH Industry Forum}. (2013) For promotion of {MPEG-DASH} 2013. [Online].
  Available: \url{http://dashif.org}
\BIBentrySTDinterwordspacing

\bibitem{netflix2015pertitle}
\BIBentryALTinterwordspacing
A.~Aaron, Z.~Li, M.~Manohara, D.~J. Cock, and D.~Ronca. (2015) {Per-Title}
  encode optimization. [Online]. Available:
  \url{https://medium.com/netflix-techblog/per-title-encode-optimization-7e99442b62a2}
\BIBentrySTDinterwordspacing

\bibitem{aom2018AV1}
\BIBentryALTinterwordspacing
{Alliance for Open Media}. (2018) {AV1} bitstream and decoding process
  specification. [Online]. Available:
  \url{https://aomedia.org/av1-bitstream-and-decoding-process-specification/}
\BIBentrySTDinterwordspacing

\bibitem{li2019avc}
Z.~Li, Z.~Duanmu, W.~Liu, and Z.~Wang, ``{AVC, HEVC, VP9, AVS2 or AV1? --- A}
  comparative study of state-of-the-art video encoders on {4K} videos,'' in
  \emph{International Conference on Image Analysis and Recognition}, 2019, pp.
  162--173.

\bibitem{ozer2019good}
\BIBentryALTinterwordspacing
J.~Ozer. (2019) Good news: {AV1} encoding times drop to near-reasonable levels.
  [Online]. Available:
  \url{https://www.streamingmedia.com/Articles/ReadArticle.aspx?ArticleID=130284}
\BIBentrySTDinterwordspacing

\bibitem{grossberg2004modeling}
M.~D. Grossberg and S.~K. Nayar, ``Modeling the space of camera response
  functions,'' \emph{IEEE Transactions on Pattern Analysis and Machine
  Intelligence}, vol.~26, no.~10, pp. 1272--1282, Oct. 2004.

\bibitem{applea}
\BIBentryALTinterwordspacing
Apple. (2016) Best practices for creating and deploying {HTTP} live streaming
  media for {iPhone} and {iPad}. [Online]. Available: \url{http://is.gd/LBOdpz}
\BIBentrySTDinterwordspacing

\bibitem{grafl2013combined}
\BIBentryALTinterwordspacing
G.~Michael, T.~Christian, H.~Hermann, C.~Wael, N.~Daniel, and B.~Stefano.
  (2013) Combined bitrate suggestions for multi-rate streaming of industry
  solutions. [Online]. Available: \url{http://alicante.itec.aau.at/am1.html}
\BIBentrySTDinterwordspacing

\bibitem{bt500subjective}
\BIBentryALTinterwordspacing
{International Telecommuniations Union}. (2012) {Methodology} for the
  subjective assessment of the quality of television pictures. [Online].
  Available:
  \url{https://www.itu.int/dms_pubrec/itu-r/rec/bt/R-REC-BT.500-13-201201-I!!PDF-E.pdf}
\BIBentrySTDinterwordspacing

\bibitem{sheikh2006statistical}
H.~R. Sheikh, M.~F. Sabir, and A.~C. Bovik, ``A statistical evaluation of
  recent full reference image quality assessment algorithms,'' \emph{IEEE
  Transactions on Image Processing}, vol.~15, no.~11, pp. 3440--3451, Nov.
  2006.

\bibitem{zhai2008cross}
G.~Zhai, J.~Cai, W.~Lin, X.~Yang, W.~Zhang, and M.~Etoh, ``Cross-dimensional
  perceptual quality assessment for low bit-rate videos,'' \emph{IEEE
  Transactions on Multimedia}, vol.~10, no.~7, pp. 1316--1324, Nov. 2008.

\bibitem{kreyszig1978introductory}
E.~Kreyszig, \emph{Introductory Functional Analysis with Applications}.\hskip
  1em plus 0.5em minus 0.4em\relax Wiley New York, 1978.

\bibitem{Rehman2015SSIMplus}
A.~Rehman, K.~Zeng, and Z.~Wang, ``Display device-adapted video
  {Q}uality-of-{E}xperience assessment,'' in \emph{Proceedings of SPIE}, 2015,
  pp. 939\,406.1--939\,406.11.

\bibitem{Duanmu2017QoE}
Z.~Duanmu, K.~Ma, and Z.~Wang, ``Quality-of-{E}xperience of adaptive video
  streaming: {E}xploring the space of adaptations,'' in \emph{ACM International
  Conference on Multimedia}, 2017, pp. 1752--1760.

\bibitem{wang2004image}
Z.~Wang, A.~Bovik, H.~Sheikh, and E.~Simoncelli, ``Image quality assessment:
  {F}rom error visibility to structural similarity,'' \emph{IEEE Transactions
  on Image Processing}, vol.~13, no.~4, pp. 600--612, Apr. 2004.

\bibitem{wang2003multiscale}
Z.~{Wang}, E.~P. {Simoncelli}, and A.~C. {Bovik}, ``Multiscale structural
  similarity for image quality assessment,'' in \emph{The 37th Asilomar
  Conference on Signals, Systems Computers}, 2003, pp. 1398--1402.

\bibitem{stellato2017osqp}
B.~Stellato, G.~Banjac, P.~Goulart, A.~Bemporad, and S.~Boyd, ``{OSQP}: {A}n
  operator splitting solver for quadratic programs,'' \emph{arXiv preprint
  arXiv:1711.08013}, Nov. 2017.

\bibitem{li2016VMAF}
\BIBentryALTinterwordspacing
Z.~Li, A.~Aaron, L.~Katsavounidis, A.~Moorthy, and M.~Manohara. (2016) Toward a
  practical perceptual video quality metric. [Online]. Available:
  \url{http://techblog.netflix.com/2016/06/toward-practical-perceptual-video.html}
\BIBentrySTDinterwordspacing

\bibitem{ma2017dipiq}
K.~{Ma}, W.~{Liu}, T.~{Liu}, Z.~{Wang}, and D.~{Tao}, ``{dipIQ}: Blind image
  quality assessment by learning-to-rank discriminable image pairs,''
  \emph{IEEE Transactions on Image Processing}, vol.~26, no.~8, pp. 3951--3964,
  Aug. 2017.

\bibitem{gao2015learning}
F.~Gao, D.~Tao, X.~Gao, and X.~Li, ``Learning to rank for blind image quality
  assessment,'' \emph{IEEE Transactions on Neural Networks and Learning
  Systems}, vol.~26, no.~10, pp. 2275--2290, Oct. 2015.

\bibitem{ffmpegH264}
\BIBentryALTinterwordspacing
{FFmpeg team}. (2018) {FFmpeg} v.2.8.15. [Online]. Available:
  \url{https://trac.ffmpeg.org/wiki/Encode/H264}
\BIBentrySTDinterwordspacing

\bibitem{ffmpegVP9}
\BIBentryALTinterwordspacing
------. (2018) {FFmpeg} v.2.8.15. [Online]. Available:
  \url{https://trac.ffmpeg.org/wiki/Encode/VP9}
\BIBentrySTDinterwordspacing

\bibitem{johnstone2004sparse}
\BIBentryALTinterwordspacing
I.~M. Johnstone and A.~Y. Lu. (2004) Sparse principal components analysis.
  [Online]. Available:
  \url{http://statweb.stanford.edu/~imj/WEBLIST/AsYetUnpub/sparse.pdf}
\BIBentrySTDinterwordspacing

\bibitem{kouzoupis2018recent}
D.~Kouzoupis, G.~Frison, A.~Zanelli, and M.~Diehl, ``Recent advances in
  quadratic programming algorithms for nonlinear model predictive control,''
  \emph{Vietnam Journal of Mathematics}, vol.~46, no.~4, pp. 863--882, Dec.
  2018.

\bibitem{hanhart2014calculation}
P.~Hanhart and T.~Ebrahimi, ``Calculation of average coding efficiency based on
  subjective quality scores,'' \emph{Journal of Visual Communication and Image
  Representation}, vol.~25, no.~3, pp. 555--564, Apr. 2014.

\bibitem{potra2000interior}
F.~A. Potra and S.~J. Wright, ``Interior-point methods,'' \emph{Journal of
  Computational and Applied Mathematics}, vol. 124, no.~1, pp. 281--302, Dec.
  2000.

\bibitem{zern2010webm}
\BIBentryALTinterwordspacing
J.~Zern and J.~Bankoski. (2010) {WebM} contributor's guide. [Online].
  Available:
  \url{https://chromium.googlesource.com/webm/contributor-guide/+/master/scripts/visual_metrics.py}
\BIBentrySTDinterwordspacing

\bibitem{thekerneltrip2018computational}
\BIBentryALTinterwordspacing
{RUser4512}. (2018) Computational complexity of machine learning algorithms.
  [Online]. Available:
  \url{https://www.thekerneltrip.com/machine/learning/computational-complexity-learning-algorithms/}
\BIBentrySTDinterwordspacing

\bibitem{ohm2012hevc}
J.~{Ohm}, G.~J. {Sullivan}, H.~{Schwarz}, T.~K. {Tan}, and T.~{Wiegand},
  ``Comparison of the coding efficiency of video coding standards—{Including}
  high efficiency video coding {(HEVC)},'' \emph{IEEE Transactions on Circuits
  and Systems for Video Technology}, vol.~22, no.~12, pp. 1669--1684, Dec.
  2012.

\bibitem{cock2016comparison}
J.~D. Cock, A.~Mavlankar, A.~Moorthy, and A.~Aaron, ``{A large-scale video
  codec comparison of x264, x265 and libvpx for practical VOD applications},''
  in \emph{Applications of Digital Image Processing XXXIX}, vol. 9971.\hskip
  1em plus 0.5em minus 0.4em\relax International Society for Optics and
  Photonics, 2016, pp. 363 -- 379.

\bibitem{tan2016video}
T.~K. {Tan}, R.~{Weerakkody}, M.~{Mrak}, N.~{Ramzan}, V.~{Baroncini}, J.~{Ohm},
  and G.~J. {Sullivan}, ``Video quality evaluation methodology and verification
  testing of {HEVC} compression performance,'' \emph{IEEE Transactions on
  Circuits and Systems for Video Technology}, vol.~26, no.~1, pp. 76--90, Jan.
  2016.

\bibitem{grois2017performance}
D.~Grois, T.~Nguyen, and D.~Marpe, ``{Performance comparison of {AV1, JEM, VP9,
  and HEVC} encoders},'' in \emph{Proceedings of SPIE Optical
  Engineering+Applications}, 2017, pp. 68--79.

\bibitem{akyazi2018comparison}
P.~{Akyazi} and T.~{Ebrahimi}, ``Comparison of compression efficiency between
  {HEVC/H.265, VP9 and AV1} based on subjective quality assessments,'' in
  \emph{IEEE International Conference on Quality of Multimedia Experience},
  2018, pp. 1--6.

\bibitem{stream2015dash}
\BIBentryALTinterwordspacing
Streamroot. (2015) How to encode multi-bitrate videos in {MPEG-DASH} for {MSE}
  based media players. [Online]. Available:
  \url{https://blog.streamroot.io/encode-multi-bitrate-videos-mpeg-dash-mse-based-media-players/}
\BIBentrySTDinterwordspacing

\bibitem{bishop2006pattern}
C.~Bishop, \emph{Pattern Recognition and Machine Learning}.\hskip 1em plus
  0.5em minus 0.4em\relax Springer-Verlag, 2006.

\end{thebibliography}

\end{document}